\documentclass[review, preprint,12pt]{elsarticle} %

\usepackage{amssymb}
\usepackage{amsmath,amsfonts,amssymb,bm,accents}
\usepackage{hyperref}
\usepackage{mathrsfs}
\usepackage{graphicx}
\usepackage{epstopdf}
\usepackage{float}
\usepackage{caption}
\usepackage{subcaption}
\usepackage{bbm}
\usepackage{mathrsfs}
\usepackage{cleveref}
\usepackage{soul}
\usepackage{accents}
\usepackage{graphicx}
\usepackage[export]{adjustbox}
\usepackage{xcolor}
\usepackage{courier} 
\usepackage{listings} 
\usepackage{tabu} 
\usepackage{booktabs}
\usepackage{longtable}
\usepackage{changepage} 
\usepackage[margin=2cm]{geometry}
\biboptions{sort&compress} 
\usepackage[section]{placeins}
\usepackage[font={small}]{caption}

\usepackage{siunitx}
\usepackage[export]{adjustbox}

\newcommand{\beq}{\begin{equation}}
\newcommand{\eeq}{\end{equation}}
\newcommand{\bea}{\begin{eqnarray}}
\newcommand{\eea}{\end{eqnarray}}

\newcommand{\dd}{\,{\rm d}}

\journal{Cement and Concrete Research}

\makeatletter
\def\@author#1{\g@addto@macro\elsauthors{\normalsize%
    \def\baselinestretch{1}%
    \upshape\authorsep#1\unskip\textsuperscript{%
      \ifx\@fnmark\@empty\else\unskip\sep\@fnmark\let\sep=,\fi
      \ifx\@corref\@empty\else\unskip\sep\@corref\let\sep=,\fi
      }%
    \def\authorsep{\unskip,\space}%
    \global\let\@fnmark\@empty
    \global\let\@corref\@empty  
    \global\let\sep\@empty}%
    \@eadauthor={#1}
}
\makeatother

\begin{document}

\begin{frontmatter}



\title{Unravelling the interplay between steel rebar corrosion rate and corrosion-induced cracking of reinforced concrete}


\author[IC]{Ev\v{z}en Korec}

\author[CTU]{Milan Jir\'{a}sek}

\author[IC]{Hong S. Wong}

\author[Oxf,IC]{Emilio Mart\'{\i}nez-Pa\~neda\corref{cor1}}
\ead{emilio.martinez-paneda@eng.ox.ac.uk}

\address[IC]{Department of Civil and Environmental Engineering, Imperial College London, London SW7 2AZ, UK}

\address[CTU]{Department of Mechanics, Faculty of Civil Engineering, Czech Technical University in Prague, Th\'{a}kurova 7, Prague - 6, 166 29, Czech Republic}

\address[Oxf]{Department of Engineering Science, University of Oxford, Oxford OX1 3PJ, UK}


\begin{abstract}
Accelerated impressed current testing is the most common experimental method for assessing the susceptibility to corrosion-induced cracking, the most prominent challenge to the durability of reinforced concrete structures. Although it is well known that accelerated impressed current tests lead to slower propagation of cracks (with respect to corrosion penetration) than in natural conditions, which results in overestimations of the delamination/spalling time, the origins of this phenomenon have puzzled researchers for more than a quarter of a century. In view of recent experimental findings, it is postulated that the phenomenon can be attributed to the variability of rust composition and density,
specifically to the variable ratio of the mass fractions of iron oxide and iron hydroxide-oxide,
which is affected by the magnitude of the applied corrosion current density. Based on this hypothesis, a corrosion-induced cracking model for virtual impressed-current testing is presented. The simulation results obtained with the proposed model are validated against experimental data, showing good agreement. Importantly, the model can predict corrosion-induced cracking under natural conditions and thus allows for the calculation of a newly proposed crack width slope correction factor, which extrapolates the surface crack width measured during accelerated impressed current tests to corrosion in natural conditions. \\ 
\end{abstract}


\end{frontmatter}


\section{Introduction}
\label{Introduction}

Corrosion is responsible for premature deterioration of 70-90\% of all reinforced concrete structures \cite{Gehlen2011-za, British_Cement_Association_BCA1997-jj}. One of the most common degradation mechanisms is corrosion-induced cracking, which takes years or decades under natural conditions. For this reason, researchers have strived for decades to accurately simulate and predict corrosion-induced cracking \cite{Angst2018a}. 

Currently, there are two methods employed to experimentally simulate corrosion-induced cracking in laboratory conditions \cite{Yuan2007}. The first option is to expose the concrete specimen to an exaggerated climate, i.e., to various intensities of contact with aggressive species such as chlorides or carbon dioxide, and possibly also to variable humidity (wetting-drying cycles). This accelerates the corrosion process to a certain extent. However, accelerated testing can still require months or years to achieve significant corrosion-induced cracking. Also, corrosion current density, steel mass loss and anodic area are evolving dynamically with the penetration of aggressive species. These are not known in advance and need to be measured in other ways, which often requires many samples to be tested at different stages of the corrosion process.


The other method to simulate corrosion-induced cracking experimentally is the impressed current technique. In this case, an electric potential is applied on the steel rebar which serves as the anode, with typically another stainless steel rebar or mesh serving as the cathode. This method results in nearly uniform corrosion and allows to accurately control corrosion current density and mass loss. At the same time, it allows for applying high corrosion current density in hundreds or thousands of \unit{\micro\ampere\per\centi\metre^2},  which drastically shortens the time of the experiment to days or hours. For all their benefits and simplicity, impressed current techniques have become very popular \cite{Angst2018a}. However, a major downside is that while corrosion current densities from 100 to 2000 \unit{\micro\ampere\per\centi\metre^2} are commonly applied in impressed current tests \cite{Angst2018a}, natural corrosion current densities are much smaller, typically about 1 \unit{\micro\ampere\per\centi\metre^2} \cite{Otieno2012a, Otieno2016a, Andrade2023, Walsh2016, andrade2023role}. This raises the question of how accurately can the impressed current technique reproduce the corrosion process and corrosion-induced cracking under natural conditions.
  
To answer this question, \citet{Alonso1996} measured the surface crack width on the impressed current test with varying applied corrosion current density and found out that lower current density resulted in faster crack propagation with respect to corrosion penetration (i.e. the thickness of the corroded steel layer). The measured differences were significant: \citet{Alonso1996} reported that the slope of linearly fitted crack width/corrosion penetration curve for the corrosion rate of 10 \unit{\micro\ampere\per\centi\metre^2} was even six times larger than for the corrosion rate of 100 \unit{\micro\ampere\per\centi\metre^2}. The decreasing trend of the slope of linearly fitted crack width/corrosion penetration curves with increasing applied corrosion current density was  
confirmed by the comprehensive experimental study of \citet{Pedrosa2017}, who found out that the slope is proportional to a negative power of the corrosion current density. Slower propagation of cracks under an increased applied corrosion current density was also observed by \citet{Vu2005,saifullah1994effect} and \citet{Mullard2011} up to approximately 200 \unit{\micro\ampere\per\centi\metre^2}. By fitting the experimental results, \citet{Vu2005} proposed a loading correction factor dependent on the applied corrosion current density to allow for the extrapolation of the time to cracking measured within accelerated tests to corrosion in natural conditions. It has to be mentioned that for very high currents over 200 \unit{\micro\ampere\per\centi\metre^2}, some authors reported an increased speed of crack propagation compared to current densities lower than 200 \unit{\micro\ampere\per\centi\metre^2} \cite{ElMaaddawy2003, Mangat1999, saifullah1994effect}. \citet{ElMaaddawy2003} suggested that for a higher corrosion rate, corrosion products have less time to be transported into the concrete pore space and a larger portion of them is thus trapped at the steel-concrete interface, which increases the induced pressure. 

Even though the decrease of crack propagation with increasing applied corrosion current up to at least 200 \unit{\micro\ampere\per\centi\metre^2} is well experimentally documented, there has been no agreement on the origins of this phenomenon. \citet{Alonso1996} and \citet{Pedrosa2017} proposed an alternative explanation, suggesting that the loading rate-dependent material properties of concrete may be responsible. The idea is that the increased strain rate would affect the fracture-related properties such that the tensile strength would increase and delay the cracking process. If this was true, given the range of rates relevant to the corrosion process, the rate-dependency of material properties would have to be caused by creep. Although creep is likely to affect the results of long-term tests conducted under natural-like corrosion current densities, the study of \citet{Aldellaa2022a} found out that creep alone does not explain the dependence of the speed of the crack propagation (with respect to corrosion penetration) on the corrosion rate. 

Recently, a new plausible explanation emerged from the study of \citet{Zhang2019c} who performed X-ray diffraction (XRD) measurements on the series of rust samples from impressed current tests conducted with different applied corrosion current densities. \citet{Zhang2019c} found out that the composition of rust, specifically the mass fractions of two main components of rust---iron oxides (mainly w{\"u}stite $\mathrm{FeO}$, maghemite $\left(\gamma-\mathrm{Fe}_2 \mathrm{O}_3\right)$ and magnetite $\left(\mathrm{Fe}_3 \mathrm{O}_4\right)$) and iron hydroxy-oxides (mainly goethite $\alpha$-FeOOH, akageneite $\beta$-FeOOH, lepidocrocite $\gamma$-FeOOH)---changed with the applied magnitude of corrosion current. This indicates that as the corrosion current density increases, the mass fraction of iron hydroxy-oxides decreases. Because iron hydroxy-oxides have a smaller density than iron oxides, the decrease in hydroxy-oxide content results in lower corrosion-induced pressures and thus slower crack evolution. Although the study of \citet{Liu2020a} identified high hydroxy-oxide content even when the corrosion current density was raised to 100 \unit{\micro\ampere\per\centi\metre^2}, the conclusions of \citet{Zhang2019c} are supported by a number of other experimental studies. \citet{Yuan2007} observed different colours (indicating different chemical compositions) of rust produced from accelerated impressed current tests than those corroded in an artificial climate environment closely simulating the natural corrosion process. \citet{Chitty2005} investigated rust from ancient Gallo--Roman reinforced concrete artefacts naturally corroding for roughly two millennia and identified mostly hydroxy-oxide goethite with marblings of iron oxides maghemite and magnetite. The analyses of naturally carbonated corrosion products in carbonated reinforced concrete facade panels by \citet{Kolio2015} identified mostly hydroxy-oxides goethite, feroxyhite ($\delta-\mathrm{FeOOH}$) and lepidocrocite with some rare findings of iron oxides magnetite and maghemite. A very high content (more than 56$\%$ mass fraction) of iron oxides in four impressed current tests accelerated to 100 \unit{\micro\ampere\per\centi\metre^2} was measured by \citet{Zhang2017}.   

These recent findings are adopted to build our hypothesis and develop a corrosion-induced cracking model capable of resolving (i) the reactive transport of dissolved iron species released from the steel surface and their precipitation into the dense rust layer adjacent to the corroding steel surface and the concrete pore space, (ii) the change of the rust chemical composition, specifically the ratio of mass fractions of iron oxides and iron hydroxy-oxides, and thus of the density of rust with the applied magnitude of corrosion current density, (iii) the corrosion-induced pressure of corrosion precipitates accumulating in the dense rust layer and the concrete pore space of concrete, and (iv) the corrosion-induced fracture of concrete, as simulated with a quasi-brittle phase-field fracture model. For the first time, the analytically derived estimates are presented: for (I) the corrosion-induced pressure of the dense layer of corrosion products, which takes into account their compressibility, and (II) the $\mathrm{Fe}^{2+}$ ions flux reduction factor determining the ratio of corrosion products precipitating in the dense rust layer and those transported further into the concrete pore space. Thus, a comprehensive model capable of simulating impressed current tests is provided. Also, the model allows to predict a newly proposed crack width slope correction factor with which the surface crack obtained from impressed current tests can be extrapolated to corrosion in natural conditions.

\section{Theory and computational framework}
\label{Sec:Theory}
The formulation of the corrosion-induced cracking model comprises several parts. Firstly, the model for reactive transport and precipitation of dissolved iron released from the corroding steel surface is described in Section \ref{subSec:ReTransMod}, allowing for the prediction of the distribution of iron oxide and iron hydroxy-oxide rust. This is followed by the discussion of the phase-field fracture model in Section \ref{SubSec:fractureWu}. Corrosion-induced mechanical stress arises from the accumulation of rust in the form of a dense rust layer (Section \ref{sec:mechanicsRustLay}) and in the pore space of concrete (Section \ref{sec:mechanics_eigenstrain}). The damage-dependent diffusivity tensor allowing for the consideration of the role of crack in the reactive transport of dissolved iron species is introduced in Section \ref{sec:DamDepTen}. The formulation of the model is concluded in Section \ref{Sec:govEq} with the summary of governing equations and related boundary conditions. \\ 
\noindent \emph{Notation}. Scalar quantities are represented by light-faced italic letters, e.g. $\phi$, Cartesian vectors by upright bold letters, e.g. $\mathbf{u}$, and Cartesian second- and higher-order tensors by bold italic letters, e.g. $\bm{\sigma}$. The symbol $ \bm{1} $ denotes the second-order identity tensor and $ \bm{I} $ represents the fourth-order identity tensor. Inner products are denoted by a number of vertically stacked dots, where the number of dots corresponds to the number of indices over which summation takes place, e.g.\ $ \bm{\sigma}:\bm{\varepsilon} = \sigma_{ij} \varepsilon_{ij}$. Finally, $ \bm{\nabla} $ and $ \bm{\nabla} \cdot $ respectively denote the gradient and divergence operators.

\FloatBarrier
\subsection{Reactive transport model}
\label{subSec:ReTransMod}
\subsubsection{Solution domain, representative volume element (RVE) and primary unknown variables}
\begin{figure}[htp]
\begin{center}
\includegraphics[scale=0.5]{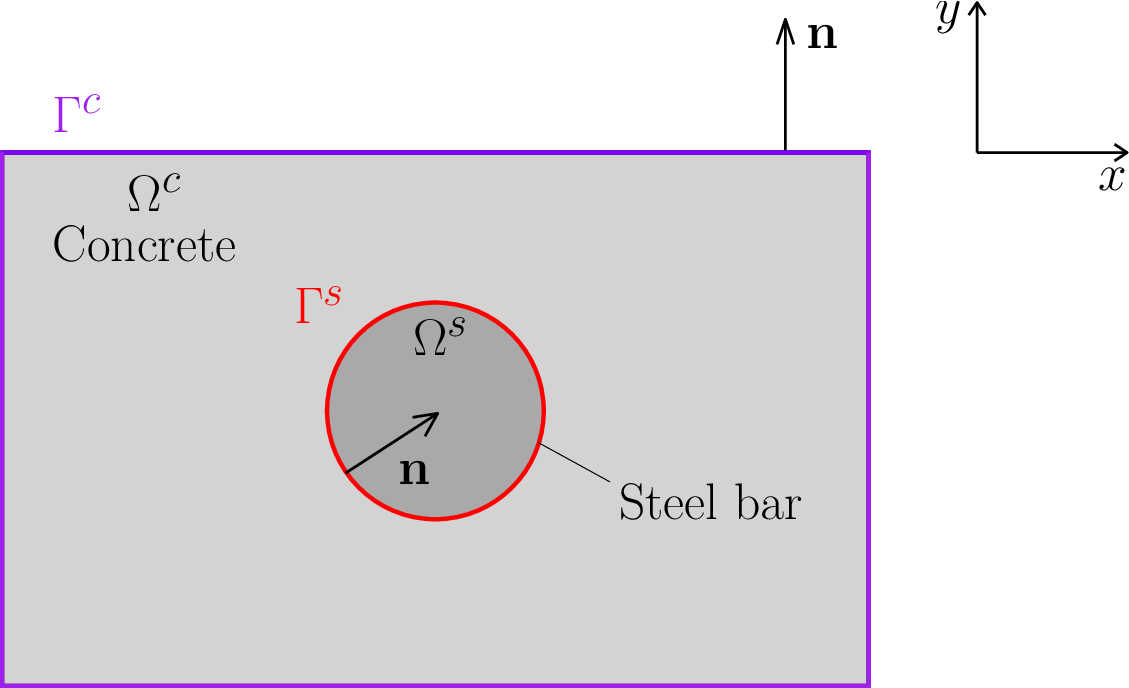}
\caption{Spatial domains and their boundaries}
\label{FigDomainBound}
\end{center} 
\end{figure}

\begin{figure}[htp]
\begin{center}
\includegraphics[scale=1.0]{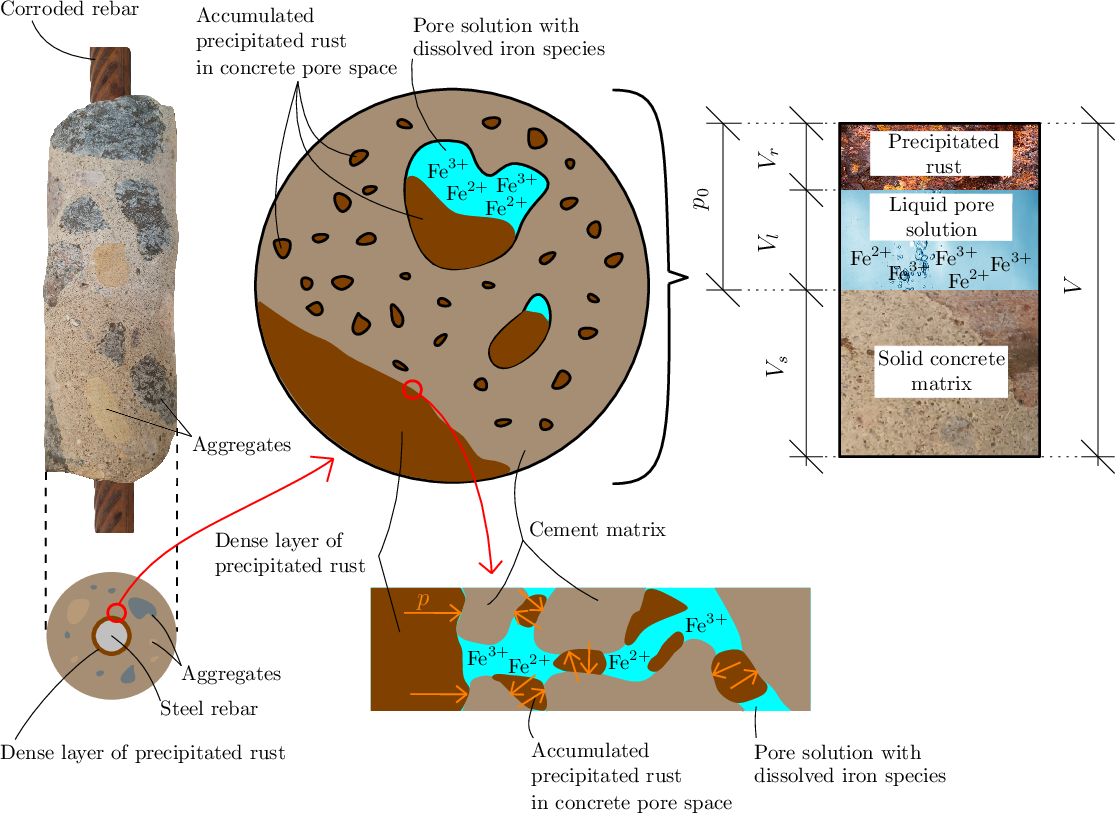}
\caption{Schematic illustration of a Representative Volume Element (RVE) of concrete at the vicinity of corroding steel, highlighting the relevant phases.}
\label{Fig2}
\end{center} 
\end{figure}

The problem is solved on a domain $ \Omega = \Omega^{c} \cup \Omega^{s} $, which comprises the concrete domain $ \Omega^{c} $ and the embedded steel domain $\Omega^{s}$ (see Fig. \ref{FigDomainBound}). The outer boundary of $\Omega$ and $\Omega^{c}$ is $\Gamma^{c}$ and the inner boundary of $ \Omega^{c} $ between the steel and concrete domains is $\Gamma^{s}$. The symbol $\mathbf{n}$ denotes the outer normal vector to the boundary $\Gamma^{c} \cup \Gamma^{s}$ of the concrete domain.   
Concrete is a porous material and within the reactive transport model, its internal structure is described by the representative volume element (RVE) depicted in Fig. \ref{Fig2}. The concrete RVE of volume $V$ is assumed to comprise the solid concrete matrix of volume $ V_s $ and the pore space of volume $V-V_s$, characterized by capillary porosity $ p_{0} = (V-V_{s})/V $, with all these quantities remaining unchanged in time. The pore space is simplified to be fully saturated with water, which is a reasonable assumption in the vicinity of the steel rebar, especially for corrosion induced by impressed current or chloride ingress. There, the pore space is gradually filled with precipitated rust, and the volume of liquid pore solution $V_{l}$ decreases to the benefit of the increasing volume of precipitated rust $V_{r}$. Volume fractions $\theta_{l} = V_{l}/V$, $\theta_{r} = V_{r}/V$ and saturation ratios $ S_{l} = \theta_{l}/p_{0} $, $ S_{r} = \theta_{r}/p_{0} $ of the liquid pore solution and the precipitated rust provide useful indicators of their current volume content. The primary unknown variables of the reactive transport problem are $ c_{II} $ and $ c_{III} $, in mols per cubic meter of the liquid pore solution.    

\FloatBarrier
\subsubsection{Distribution and chemical composition of rust in concrete}
Microscopy investigations of corroded reinforced concrete samples \cite{Chitty2005,Care2008,Wong2010a} show that a portion of the rust accumulates in the immediate vicinity of the corroded steel rebar as a dense rust layer while the remaining rust is located in the concrete pore space up to a certain distance from the rebar (see Fig. \ref{Fig2}). 

In samples corroded under both natural and accelerated conditions, the main corrosion products found with X-ray diffraction (XRD) measurements were iron oxides (w{\"u}stite $\mathrm{FeO}$, hematite $\alpha-\mathrm{Fe}_2 \mathrm{O}_3$, magnetite $\mathrm{Fe}_3 \mathrm{O}_4$) and iron hydroxy-oxides (goethite $\alpha-\mathrm{FeO(OH)}$, akaganeite $\beta-\mathrm{FeO(OH)}$, lepidocrocite $\gamma-\mathrm{FeO(OH)}$, feroxyhyte $\delta-\mathrm{FeOOH}$) \cite{Chitty2005, Zhao2011b, Zhang2019c}. It should be noted that the XRD method can determine only the presence of crystalline materials, but some corrosion products are known to be amorphous \cite{Zhao2011b}.   
\begin{figure}[!htb]
    \centering
    \includegraphics[width=0.5\textwidth]{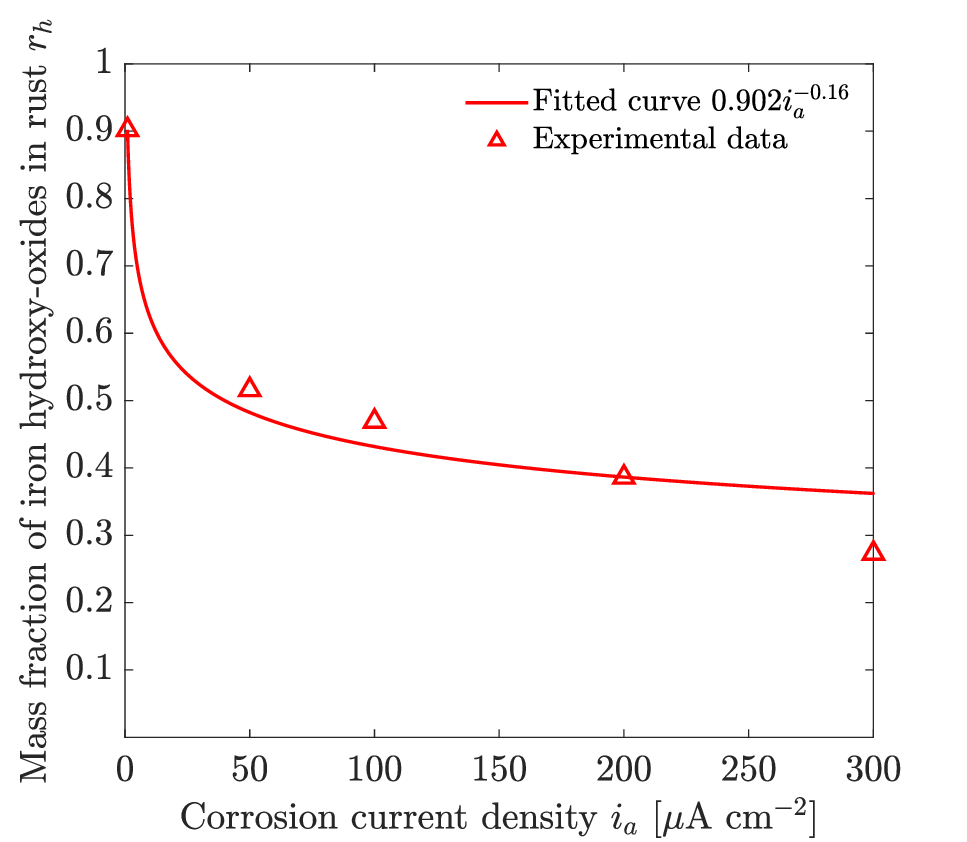}
    \caption{The dependence of the mass fraction of hydroxy-oxides rust $r_h$ on corrosion current density. $r_h$ decreases dramatically when accelerated from the typical range of natural corrosion current densities about 1 \unit{\micro\ampere\per\centi\metre^2}. Data adopted from \citet{Zhang2019c}.}
\label{massRatRust}
\end{figure}
One of the critical inputs of the model is the dependence of the mass fraction of iron hydroxy-oxides in rust on applied corrosion current density because this determines the density of produced rust and thus the induced pressure. This curve is obtained by fitting the experimentally measured data of \citet{Zhang2019c} with a power function $k_{1}(i_{a}/i_{a,ref})^{k_{2}} $ (see Fig. \ref{massRatRust}), where $i_{a,ref} = 1 $ \unit{\micro\ampere\per\centi\metre^2} is the reference corrosion current density. The first measurement of \citet{Zhang2019c} came from a naturally corroded sample for which the value of corrosion current density was not documented and thus it was estimated as 1 \unit{\micro\ampere\per\centi\metre^2}, which is a typical value measured during the natural corrosion process in reinforced concrete \cite{Otieno2012a, Otieno2016a, Andrade2023, Walsh2016, andrade2023role}. Even though experimental measurements of the mass ratio of particular corrosion products are very scarce in the current literature, the results of \citet{Zhang2019c} agree with the studies on samples corroded under natural conditions \cite{Chitty2005,Kolio2015} and with accelerated impressed current tests \cite{Zhang2017}. In Fig. \ref{massRatRust}one
can see that $r_h$ drops dramatically from 0.9 for 1 \unit{\micro\ampere\per\centi\metre^2} to roughly 0.5 for 50 \unit{\micro\ampere\per\centi\metre^2}, while a subsequent decrease is significantly more moderate.  

\FloatBarrier
\subsubsection{Iron ion transport and precipitation}
\begin{figure}[htp]
\begin{center}
\includegraphics[scale=0.9]{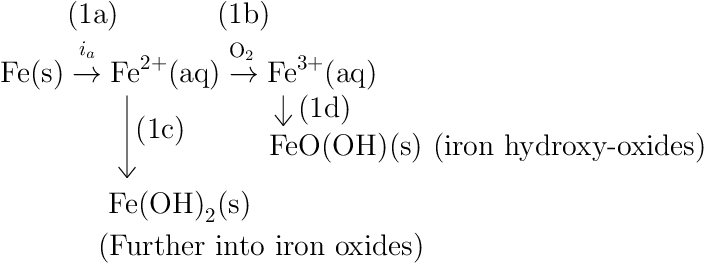}
\caption{Scheme of considered chemical reactions}
\label{FigChemScheme}
\end{center} 
\end{figure}

An alkaline concrete pore solution allows for the formation of a nanometre-thick protective semiconductive layer on the steel rebar surface, which significantly impedes the corrosion process. However, this layer can be disrupted, typically by chlorides or due to the carbonation of concrete. If water and oxygen are present, corrosion can initiate. A corrosion current of density $i_a$ starts to flow between anodic and cathodic regions and $ \mathrm{Fe}^{2+} $ ions are released from the steel surface into the pore solution by reaction (\ref{reac1}). The dissolved ferrous $ \mathrm{Fe}^{2+} $ ions then undergo a series of complex chemical reactions forming a number of intermediary products, see for instance \cite{Furcas2022,Wieland2023} for more details. For the purpose of mechanical modelling, the complexity of the chemical reaction system is reduced to three arguably critical reactions (see Fig. \ref{FigChemScheme}). The first considered option for $ \mathrm{Fe}^{2+} $ ions is to be oxidised to $\mathrm{Fe}^{3+}$ ions (reaction (\ref{reac2})), which then further precipitate into iron hydroxy-oxides ($\mathrm{FeO(OH)}$); see reaction (\ref{reac3}). The alternative option to this process is direct precipitation of $ \mathrm{Fe}^{2+} $ ions into $\mathrm{Fe(OH)_{2}}$, which is assumed to be further transformed into iron oxides, especially into mixed $\mathrm{Fe}^{2+}$--$\mathrm{Fe}^{3+}$ oxide $\mathrm{Fe_{3}O_{4}}$ and $\mathrm{Fe_{2}O_{3}}$ \cite{ZHAO201619}, though this transformation is not explicitly simulated.
\begin{subequations}
\begin{align}
\label{reac1}
& 2 \mathrm{Fe}+\mathrm{O}_2+2 \mathrm{H}_2 \mathrm{O} \rightarrow 2 \mathrm{Fe}^{2+}+4 \mathrm{OH}^{-} \\
\label{reac2}
& 4 \mathrm{Fe}^{2+}+\mathrm{O}_2+2 \mathrm{H}_2 \mathrm{O} \rightarrow 4 \mathrm{Fe}^{3+}+4 \mathrm{OH}^{-} \\
\label{reac3}
& \mathrm{Fe}^{2+}+2 \mathrm{OH}^{-} \rightarrow \mathrm{Fe}(\mathrm{OH})_2 \\
\label{reac4}
& \mathrm{Fe}^{3+}+3 \mathrm{OH}^{-} \rightarrow \mathrm{FeO}(\mathrm{OH})+\mathrm{H}_2 \mathrm{O}
\end{align}
\end{subequations}
Experimental studies indicate that iron oxides $\mathrm{Fe_{2}O_{3}}$ and $\mathrm{Fe_{3}O_{4}}$ together with iron hydroxy-oxides $\mathrm{FeO(OH)}$ constitute the majority of produced rust for both accelerated impressed current tests and samples corroding in natural conditions \cite{Zhao2011b, Kolio2015, Chitty2005, Zhang2019c, Liu2020a, Sola2019a}. However, critical for this study, based on the experimental results of \citet{Zhang2019c}, it is assumed that the ratio of mass fractions of iron oxides and hydroxy-oxides changes with the corrosion current density (see Fig. \ref{massRatRust}).       
The reactive transport of $ \mathrm{Fe}^{2+} $ and $ \mathrm{Fe}^{3+} $ in $\Omega^{c}$ is described by mass-conserving diffusion equations 
\begin{equation}\label{total_mass_derivative_ionic_N_2}
\frac{\partial \left(\theta_{l}c_{II}\right)}{\partial t} - \bm{\nabla} \cdot \left(\theta_{l}\bm{D}_{II}\cdot\nabla c_{II} \right) =  \theta_{l} (-R_{II}-R_{o}) \,\,\ \text{ in } \,\, \Omega^{c}, \quad R_{II} =  k^{II \rightarrow III}_{r} c_{II}c_{ox}, \quad R_{o} = k^{II \rightarrow o}_{r} c_{II}
\end{equation}  
\begin{equation}\label{total_mass_derivative_ionic_N_3}
\frac{\partial \left(\theta_{l}c_{III}\right)}{\partial t} - \bm{\nabla} \cdot \left(\theta_{l}\bm{D}_{III}\cdot\nabla c_{III} \right) =  \theta_{l} (R_{II}-R_{h}) \,\,\ \text{ in } \,\, \Omega^{c}, \quad R_{h} = k^{III \rightarrow h}_{r} c_{III}
\end{equation} 
previously derived in \cite{Korec2023}. Small deformations are assumed, the velocity of the solid concrete matrix is neglected, and the flux term is scaled with liquid volume fraction following \citet{Marchand2016}.

In Eqs. (\ref{total_mass_derivative_ionic_N_2})--(\ref{total_mass_derivative_ionic_N_3}), $ \bm{D}_{II} $ and $ \bm{D}_{III} $ are the second-order diffusivity tensors of $ \mathrm{Fe}^{2+} $ and $ \mathrm{Fe}^{3+} $ ions. $ R_{II} $,$ R_{o} $ and $ R_{h} $ are the reaction rates of reactions (\ref{reac2}), (\ref{reac3}) and (\ref{reac4}), respectively, with $ k^{II \rightarrow III}_{r} $, $k^{II \rightarrow o}_{r}$ and $k^{III \rightarrow h}_{r}$ being the corresponding  reaction constants. Since the ratio of the mass fractions of iron oxides and hydroxy-oxides in rust changes with the corrosion current density, reaction constants $ k^{II \rightarrow o}_{r} $ and $ k^{III \rightarrow h}_{r} $ have to depend on the corrosion current density too. Previous studies provided  estimates of $ k^{II \rightarrow III}_{r} = 0.1 $ mol$^{-1}$m$^{3}$s$^{-1}$ \cite{stefanoni_kinetic_2018} and $k^{III \rightarrow h}_{r} = 2 \cdot 10^{-4}$ s$^{-1}$ \cite{Leupin2021},
but $k^{II \rightarrow o}_{r}$ [s$^{-1}$] was not known. For this reason, $k^{III \rightarrow h}_{r}$ is considered as fixed, and $k^{II \rightarrow o}_{r}$ is varied with the corrosion current density and is estimated such that the mass ratio documented in Fig. \ref{massRatRust} is achieved for a closed system with a given initial concentration of $ \mathrm{Fe}^{2+} $ ions. It can be shown that, in the limit of $ t \rightarrow \infty $, the ratio $c_{II}/c_{III}$ approaches $k^{II \rightarrow III}_{r} c_{ox}/k^{II \rightarrow o}_{r}$, from which $k^{II \rightarrow o}_{r} = k^{II \rightarrow III}_{r} c_{ox} (1-r_{h})M_{h}/M_{o}$ is estimated. $ M_{o} $ and $M_{h}$ refer to the molar masses of oxide rust and hydroxy-oxide rust respectively. These estimates are theoretically accurate only for large times, but the half-time of $ \mathrm{Fe}^{2+} $ ion transformation is only about 2900 s for $i_{a} = 1$ \unit{\micro\ampere\per\centi\metre^2} and steadily decreases to approximately 560 s for $i_{a} = 500$ \unit{\micro\ampere\per\centi\metre^2}. This is very short compared to the advance of corrosion penetration---the shortest further discussed case studies lasted over 26 days or 6 hours for the cases of $i_{a} = 1$ or $i_{a} = 500$ \unit{\micro\ampere\per\centi\metre^2}, respectively. For all calculations, the constant oxygen concentration $c_{ox} = 0.28 $ mol m$^{-3}$ is assumed as in the study of \citet{Zhang2021}. 

The volume fractions $\theta_{o}$ and $\theta_{h}$ of precipitated iron oxides and hydroxy-oxides are respectively governed by 
\begin{equation}\label{total_mass_derivative_crystals_N_2}
\frac{\partial \theta_{o}}{\partial t} =  \dfrac{M_{o}}{\rho_{o}} \theta_{l} R_{o} \,\,\ \text{ in } \,\, \Omega^{c}
\end{equation} 
\begin{equation}\label{total_mass_derivative_crystals_N_3}
\frac{\partial \theta_{h}}{\partial t} =  \dfrac{M_{h}}{\rho_{h}} \theta_{l} R_{h} \,\,\ \text{ in } \,\, \Omega^{c}
\end{equation} 
in which $ \rho_{o} $ and $\rho_{h}$ are densities of oxide rust and hydroxy-oxide rust respectively, and $ M_{h} $ and $M_{o}$ denote the molar masses of hydroxy-oxide rust and oxide rust, respectively. Immobile precipitated rust gradually fills the concrete pore space and reduces the liquid volume fraction $ \theta_l = p_0 - \theta_o - \theta_h $.  

Reactive transport equations (\ref{total_mass_derivative_ionic_N_2})--(\ref{total_mass_derivative_ionic_N_3}) require boundary conditions. On the outer concrete boundary, zero flux is assumed such that $\textbf{n} \cdot \left(\theta_{l}\bm{D}_{II}\cdot\nabla c_{II} \right) = 0$ and $ \textbf{n} \cdot \left(\theta_{l}\bm{D}_{III}\cdot\nabla c_{III} \right) = 0 $ on $\Gamma^{c}$. On the corroding steel rebar surface $ \Gamma^{s} $, the inward influx of ferrous ions follows Faraday's law        
\begin{equation}\label{FarLaw_1}
J_{II} = k_{f}\frac{i_{a}}{z_{a}F} = -\mathbf{n} \cdot \left(-\boldsymbol{D}_{II} \cdot \nabla c_{II} \right)
\end{equation}
where $ i_{a} $ is the corrosion current density, $F$ is Faraday's constant and $ z_{a} = 2 $ represents the number of electrons exchanged in anodic reaction (\ref{reac1}) per one atom of iron. Some of the released $\mathrm{{Fe}^{2+}}$ ions precipitate into a dense rust layer instead of being released further into concrete pore space. This effect is taken into consideration with the flux reduction coefficient $k_{f} \in \left\langle 0,1 \right\rangle$, which is discussed in the following section. 

\subsubsection{Flux reduction coefficient $k_{f}$}
\label{sec:chemFlux}

\begin{figure}[htp]
\begin{center}
\includegraphics[scale=1]{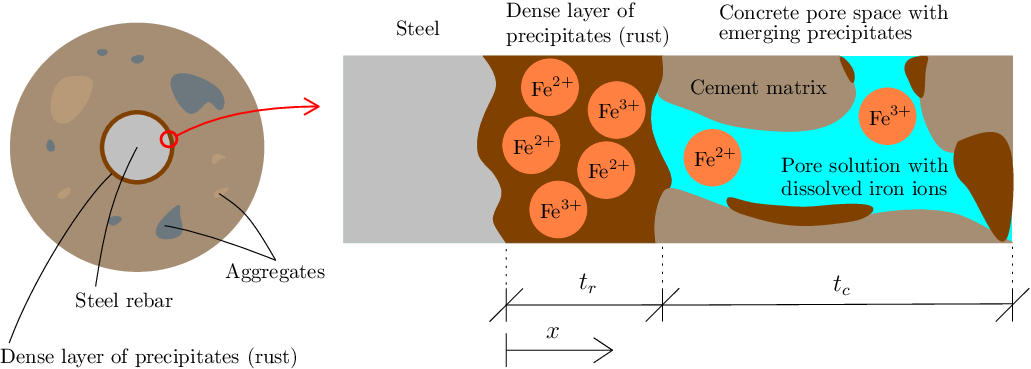}
\caption{Schematic illustration of rust precipitation in the vicinity of the steel rebar. A portion of the released $\mathrm{Fe}^{2+}$ ions and emerging $\mathrm{Fe}^{3+}$ ions precipitates to form a dense rust layer adjacent to the yet uncorroded steel surface while the remainder is transported further in concrete pore space where it eventually precipitates.}
\label{Fig3}
\end{center} 
\end{figure}

During corrosion, ferrous ions are released from the steel surface by electrochemical reaction (\ref{reac1}) and their flux follows Faraday's law $J_{II,Far} = i_{a}/(z_{a}F)$. A portion of the released $\mathrm{Fe}^{2+}$ ions and emerging $\mathrm{Fe}^{3+}$ ions precipitate to form a dense rust layer in the vicinity of the rebar while the remainder is transported further into the pore space to eventually precipitate there (see Fig. \ref{Fig3}). The thickness of the rust layer is typically much smaller compared to the rebar radius. For this reason, the rust layer is not explicitly geometrically considered during domain discretization for numerical calculation, as this would strongly exacerbate the computational requirements. Instead, the steel domain is handled as unchanged in time, and the mechanical pressure of the accumulating dense rust layer on concrete is taken into consideration by additional pressure $p$ on the steel boundary, further discussed in Section \ref{sec:mechanicsRustLay}. Thus, in reality, the flux of ferrous ions prescribed on the steel boundary is only the portion of this flux that did not precipitate into the dense rust layer. This flux reduction is mathematically incorporated by the flux reduction coefficient $k_{f}$ multiplying Faraday's law-dictated flux $J_{II,Far}$. 

To estimate $k_{f}$, let us assume instantaneous precipitation of ferric ions and constant given liquid volume fraction $\theta_{l}$, and let us simplify reactive transport equation (\ref{total_mass_derivative_ionic_N_2}) into its one-dimensional stationary form  
\begin{subequations}\label{statDiffEq}
\begin{align}
\label{statDiffEqRust}
D_{r,II}\dfrac{\dd^{2} c_{II}}{\dd x^{2}} &= k^{II \rightarrow o}_{r} c_{II} + k^{II \rightarrow III}_{r} c_{II}c_{ox} \,\,\,\,\,\,\, \text{ in } \,\, x \in \left\langle 0,t_{r} \right)  \\ 
\label{statDiffEqCon}
D_{c,II}\dfrac{\dd^{2} c_{II}}{\dd x^{2}} &= k^{II \rightarrow o}_{r} c_{II} + k^{II \rightarrow III}_{r} c_{II}c_{ox} \,\,\,\,\,\,\, \text{ in } \,\, x \in \left\langle t_{r}, t_{r}+t_{c} \right\rangle
\end{align}
\end{subequations}
Equation (\ref{statDiffEq}) is solved on an evolving domain $\left\langle 0,t_{r}+t_{c} \right\rangle$, which is separated into the dense rust layer subdomain $\left\langle 0,t_{r} \right) $ with diffusivity $D_{r,II}$ and  the concrete subdomain $\left\langle t_{r},t_{r}+t_{c} \right) $ with diffusivity $D_{c,II}$, where $ t_{r} $ changes in time (see Fig. \ref{Fig3}). For the sake of simplicity, the thickness of the dense rust layer is considered to be equal to the thickness of the corroded steel layer predicted by Faraday's law. 

It is possible to solve equations (\ref{statDiffEqRust}) and (\ref{statDiffEqCon}) independently and analytically, considering Faraday's law dictated flux $J_{II,Far}$ at $x = 0$, $ c_{II} = 0 $ at $x = t_{r}+t_{c} $ and a given unknown flux value at the current interface $ x = t_{r} $. From the assumption of flux continuity at $x=t_{r}$, a solution for $ k_{f} $ is obtained. If $ D_{c,II} $ is further replaced by $ S_{l}D_{c,II} $ to consider the effect of pore-clogging with precipitates on $ \mathrm{Fe}^{2+} $ diffusivity, the resulting flux can be expressed as
\begin{equation}
J_{II} = k_{f} \frac{i_{a}}{z_{a}F}
\end{equation}
where the flux reduction coefficient
\begin{equation}
k_{f} = \dfrac{2{\rm e}^{
A_{r}} \sqrt{S_{l}D_{c,II}} \coth A_c}{\left(1+{\rm e}^{2A_{r}}\right)\left(\sqrt{S_{l}D_{c,II}} \coth A_c + \sqrt{D_{r,II}} \tanh A_r\right)}
\end{equation}
is calculated using rust-related and concrete-related constants 
\begin{equation}
A_{r} = t_{r}\sqrt{\frac{k^{II \rightarrow o}_{r}+c_{ox} k^{II \rightarrow III}_{r}}{D_{r,II}}}, \quad A_{c} = t_{c}\sqrt{\frac{k^{II \rightarrow o}_{r}+c_{ox} k^{II \rightarrow III}_{r}}{S_{l}D_{c,II}}}
\end{equation}
Based on the results of our previous study \cite{Korec2023}, it is set $ t_{c} = 2 $ mm.    

\FloatBarrier
\subsection{Phase-field description of precipitation-induced cracks}
\label{SubSec:fractureWu}
In this model, the phase-field cohesive zone model (\texttt{PF-CZM}) of \citet{Wu2017,Wu2018} is employed, which allows for the calibration of the softening curve such that the quasi-brittle fracture nature of concrete is captured. Here a brief summary of the governing equations of the \texttt{PF-CZM} model and the necessary background is provided. For the derivation and more detailed explanation, see section 2.4 in Ref. \cite{Korec2023}. The fracture problem is solved exclusively on the concrete domain $ \Omega^{c} $ while the steel in  $ \Omega^{s} $ is assumed to remain linear elastic. The primary unknown variables in $ \Omega^{c} $ are the displacement vector 
\begin{equation}\label{setDispl}
\mathbf{u}(\mathbf{x},t) \in \mathbb{U} = \lbrace \forall t \geq 0: \mathbf{u}(\mathbf{x},t) \in W^{1,2}(\Omega)^{d}; \mathbf{u}(\mathbf{x},t) = \overline{\mathbf{u}}(\mathbf{x},t)\text{ in }\Gamma^{u} \rbrace
\end{equation} 
 and the phase-field variable 
\begin{equation}\label{setPF}
\phi(\mathbf{x},t) \in \mathbb{P} = \lbrace \forall t \geq 0: \phi(\mathbf{x},t) \in W^{1,2}(\Omega^{c}); 0 \leq \phi(\mathbf{x},t) \leq 1 \text { in } \Omega^{c}; t_{1} \leq t_{2} \Longrightarrow \phi(\textbf{x},t_{1}) \leq \phi(\textbf{x},t_{2}) \rbrace
\end{equation} 
where $ \Gamma^{u} \subset (\Gamma^{c} \cup \Gamma^{s}) $ is the portion of concrete boundary with prescribed displacement $  \overline{\mathbf{u}}$. In (\ref{setDispl}) and (\ref{setPF}), $d = 2,3$ is the geometric dimension of the problem and function space $ W^{1,2}(\Omega^{c})^{d} $ is the Cartesian product of $ d $ Sobolev spaces $ W^{1,2}(\Omega^{c}) $ consisting of functions with square-integrable weak derivatives. The phase-field variable $ \phi $ characterises the current damage of concrete such that $\phi=0$ for undamaged material and $\phi=1$ for a fully cracked material. Note that $\phi$ is not the traditional damage variable that directly reflects the relative change of stiffness, but it is linked to the material integrity through the degradation function, to be introduced later.

The small-strain tensor $\bm{\varepsilon} = \nabla_{s}\mathbf{u} = (\nabla \mathbf{u} + (\nabla \mathbf{u})^{T})/2$ is used, and the total strain $ \bm{\varepsilon} = \bm{\varepsilon}_{e} + \bm{\varepsilon}_{\star} $ is additively decomposed into the elastic part $ \bm{\varepsilon}_{e} $ and the precipitation eigenstrain $ \bm{\varepsilon}_{\star} $. Because damage limits the stress that can be carried by the material, the Cauchy stress tensor  is calculated as   
\begin{equation}\label{CauchyStressTensor2}
\bm{\sigma} = g(\phi)\mathcal{\bm{C}}_{e}^{c}:(\bm{\varepsilon} - \bm{\varepsilon}_{\star}) 
\end{equation} 
where $\mathcal{\bm{C}}_{e}^{c} $ is the fourth-order isotropic elastic stiffness tensor of concrete,
characterized by Young's modulus $E$ and Poisson's ratio $\nu_c$,
and 
$g(\phi)$ is the degradation function that describes the remaining integrity of the material. Typically, it is a function that satisfies conditions  $ g(0) = 1 $ and $ g(1) = 0$ and is non-increasing and continuously differentiable in $[0,1]$. 

Assuming tractions $ \overline{\mathbf{t}} $ prescribed on $ \Gamma^{t} = (\Gamma^{c} \cup \Gamma^{s})\setminus\Gamma^{u} $ and a volume force $\overline{\mathbf{b}} $ acting on $ \Omega^{c} $, the governing equations and the related boundary conditions for the coupled damage-displacement problem read
\begin{subequations}\label{govEqFr3}
\begin{align}
\bm{\nabla} \cdot\left(g(\phi) \mathcal{\bm{C}}_{e}^{c}: (\bm{\nabla}_{s} \mathbf{u} - \bm{\varepsilon}_{\star}\right)+\overline{\mathbf{b}} &=0 \,\,\,\,\,\,\, \text{ in } \,\, \Omega^{c} \label{govEqFr2a}\\
\mathbf{u} = \overline{\mathbf{u}}\,\,\,\,\,\,\, \text{ on } \,\,\Gamma^{u}, \quad \bm{\sigma} \cdot \mathbf{n} &= \overline{\mathbf{t}}\,\,\,\,\,\,\, \text{ on } \,\,\Gamma^{t} \label{govEqFr3a} \\ 
-\dfrac{1}{2}\frac{\dd g(\phi)}{\mathrm{~d} \phi} \mathcal{H}(t)+\dfrac{\ell}{\pi} G_{f} \nabla^{2} \phi-\dfrac{G_{f}}{\pi \ell}(1-\phi) &=0 \,\,\,\,\,\,\, \text{ in } \,\, \Omega^{c} \label{govEqFr3b} \\
\nabla \phi \cdot \mathbf{n} &= 0\,\,\,\,\,\,\, \text{ on } \,\,\Gamma^{c} \cup \Gamma^{s} 
\end{align}
\end{subequations} 
where $G_f$ is the fracture energy and $ \ell $ is the characteristic phase-field length scale governing the size of the process zone \cite{Kristensen2021}. To make the results independent of $\ell$, it is recommended to use $\ell \leq \text{min} \left( 8 \ell_{irw} / 3 \pi, \, L/100 \sim L/50 \right)$ where
 $L$ is a characteristic length of the structure (e.g., the beam depth) and $ \ell_{irw} = \widetilde{E}G_{f}/f^{2}_{t} $ is the Irwin internal length, evaluated from the fracture energy,
 tensile strength $f_t$ and  elongation modulus $ \widetilde{E} = E(1 - \nu)/((1+\nu)(1-2 \nu))$.

In the crack process zone, the characteristic size of the finite elements must be sufficiently small to provide good resolution of high strain gradients, typically 5-7 times smaller than $\ell$ \cite{Kristensen2021}. In (\ref{govEqFr3b}) the crack driving force history function 
\begin{equation}\label{historyFunc}
\mathcal{H}(t) = \displaystyle\max_{t\in[ 0, T] \rangle} \left(\widetilde{H}, H(t)\right), \quad   \widetilde{H}=\frac{f^{2}_{t}}{2 \widetilde{E}}, \quad H(t)=\frac{\left\langle \bar{\sigma}_{1}\right\rangle^{2}}{2 \widetilde{E}}
\end{equation} 
is employed to secure the irreversibility of damage \cite{Miehe2015a}. 
The function $ \mathcal{H}(t) $ stores the maximum reached value of the crack driving force $ H(t) $ in time. Crack nucleation occurs when $ H(t)$ exceeds the damage nucleation threshold $ \widetilde{H} $ calculated from the basic properties of concrete. The function $ \langle \bar{\sigma}_{1} \rangle = (\bar{\sigma}_{1} + |\bar{\sigma}_{1}|)/2 $ is the positive part of the maximum principal value of the effective stress tensor $\bar{\bm{\sigma}} = \mathcal{\bm{C}}_{e}^{c}:(\bm{\varepsilon} - \bm{\varepsilon}_{\star})$. 

The softening curve resulting from the 
phase-field
 model is affected by the choice of the degradation function $g(\phi)$. For the \texttt{PF-CZM} model, \citet{Wu2017} proposed to use 
\begin{equation}\label{damFunWu1}
g(\phi)=\frac{(1-\phi)^{p}}{(1-\phi)^{p}+a_{1} \phi (1+a_{2} \phi+a_{3}  \phi^{2})}
\end{equation}
This function is calibrated  by an appropriate choice of parameters $ p \geq 2 $, $ a_{1} > 0 $, $ a_{2} $ and $ a_{3} $, based on the required shape of the softening curve,
characterized by the ratios
\begin{equation}  
\beta_{w} = \dfrac{w_{c}}{w_{c,lin}} \quad \text{and} \quad \beta_{k} = \dfrac{k_{0}}{k_{0,lin}} 
\end{equation}
where $w_c$ is the crack opening at zero stress (full softening), $k_0$ is the initial slope of the softening curve (i.e., its negative slope at the onset of cracking), and 
\begin{equation}  
 w_{c,lin} = \dfrac{2G_{f}}{f_{t}} \quad \text{and} \quad k_{0,lin} = -\frac{f_{t}^{2}}{2G_{f}} 
\end{equation}
are the reference values that would correspond to linear softening. 

For softening laws with zero stress attained at a finite value of crack opening, the parameter $p$ is set to 2, and for laws that approach zero stress only
asymptotically, a value greater than 2 is appropriate.
The other parameters of the degradation function (\ref{damFunWu1}) are then evaluated as
\begin{equation}\label{a1a2a3}
a_{1}=\frac{4}{\pi} \frac{\ell_{irw}}{\ell}, \quad a_{2}=2 \beta_{k}^{2 / 3}-p-\frac{1}{2}, \quad a_{3}= \begin{cases} \beta_{w}^{2}/2-a_{2}-1 & \text{if} \quad p=2 \\ 0 & \text{if} \quad p>2  \end{cases}
\end{equation}    
For concrete, the experimentally measured softening can usually be reasonably represented by the 
Hordijk-Cornelissen cohesive law \cite{Cornelissen1962},
for which
\begin{equation}\label{CornelissenSoft}
w_{c}=5.1361 \frac{G_{f}}{f_{t}}, \quad k_{0}=-1.3546 \frac{f_{t}^{2}}{G_{f}}
\end{equation}  
and thus $\beta_w=5.136/2=2.568$ and $\beta_k=1.3546/0.5=2.7092$.
\citet{Wu2017} demonstrated that his version of the phase-field model indeed provides a close approximation of the Hordijk-Cornelissen curve if the parameters of the degradation function (\ref{damFunWu1})
are set to   
\begin{eqnarray}
 a_{2}&=&2 \beta_{k}^{2 / 3}-p-\frac{1}{2}
 =2 \cdot 2.7092^{2 / 3}-2-\frac{1}{2} = 1.3868
 \\
 a_{3}&=&  \beta_{w}^{2}/2-a_{2}-1 = 2.568^{2}/2-1.3868-1=0.9106
\end{eqnarray}   
These are the values adopted in the present study.
The parameter $a_1$, linked to the choice of the 
characteristic length $\ell$, will be evaluated from
the first formula in Eq. (\ref{a1a2a3}).

\FloatBarrier
\subsection{Pressure of the accumulated rust layer}
\label{sec:mechanicsRustLay}

\begin{figure}[!htb]
    \centering
    \includegraphics[width=1\textwidth]{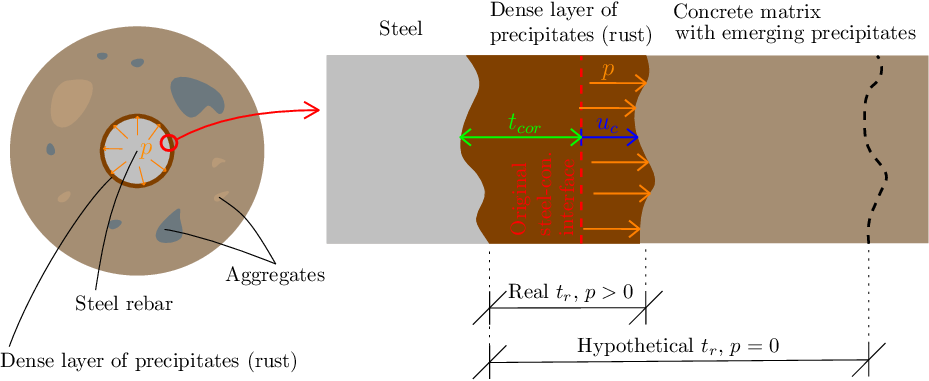}           
\caption{Schematic illustration of the pressure induced by rust precipitation in the vicinity of the rebar. Rust has a lower density than steel and the dense rust layer thus tends to occupy a larger volume (characterised by its thickness $ t_{r} $) than was vacated by the corrosion process (characterised by $ t_{cor} $). However, the rust volumetric expansion is constrained by the surrounding concrete matrix. Thus, the rust layer exerts pressure $ p $ on the concrete boundary, which leads to displacement $ u_{c} $ such that the dense rust layer occupies volume characterised by thickness $t_r=t_{cor} + u_{c}$.}
\label{fig:rusLayPresScheme}
\end{figure}

As was discussed in previous sections, $\mathrm{Fe}^{2+}$ and $\mathrm{Fe}^{3+}$ ions eventually precipitate into rust that either forms a dense rust layer in the vicinity of steel rebar or
gradually fills the concrete pore space. Rust has a significantly lower density than original iron, typically by 3 - 6 times \citep{Angst2019a}. For this reason, if the volume characterised by thickness $ t_{cor} $ is vacated by the corrosion process, the dense rust layer would occupy 3 - 6 times larger volume under unconstrained conditions (see Fig. \ref{fig:rusLayPresScheme}). However, the rust volumetric expansion is constrained by the concrete matrix and the expansion of the rust layer thus exerts pressure $ p $ on the concrete boundary. Constrained rust volumetric expansion in the pore space also causes pressure on pore walls, which is described by the precipitation eigenstrain, as explained in the following section. 

To estimate the pressure $p$ of a dense rust layer on concrete, let us virtually isolate the cylinder containing a rebar of radius $a$ and the adjacent layer of concrete with radius $b = \alpha a$ from the studied concrete specimen (see Fig. \ref{axiProblemFig}). Assuming the length of the cylinder $ L \gg \text{max}(a,b) $, isotropic linear elasticity of concrete and symmetric boundary conditions, the thick-walled cylinder can be described as an axisymmetric linear elastic plane-strain problem. If the damage is simplified to be uniformly distributed across the cylinder, the displacement on the inner concrete boundary $u_{c}$ for a given pressure $p$ reads (the derivation is provided in \ref{Sec:A2})  
\begin{equation}\label{displ4}
u_{c} =  C_{c}\, p, \quad C_{c} = \dfrac{a \left(\alpha^2 + 1-2 \nu_{c} \right)(1+\nu_{c})}{E_{c,d}(\alpha^2-1)}
\end{equation} 
where $\nu_c$ is Poisson's ratio of concrete. $E_{c,d} = g(\phi)E_{c}$ is the damaged secant modulus of concrete, calculated as the product of Young's modulus of concrete $E_{c}$ and degradation function $g(\phi)$ (see Section \ref{SubSec:fractureWu} for more details) evaluated on the inner concrete boundary. The concrete compliance constant $C_{c}$ thus encapsulates the impact of concrete sample geometry and material properties.

Both $u_{c}$ and $p$ in Eq. (\ref{displ4}) are unknown, and thus it is necessary to link them by an additional equation. For this reason, let us now analyse the relation between pressure $p$ compressing the rust layer and the volume $V_{r,d}$, which is (for small thicknesses) proportional to $ t_{r} = u_c + t_{cor}$. Let us assume that the bulk modulus of rust $K_r$ is constant and equal to $  E_r/(3(1-2\nu_r))$, with $E_r$ and $\nu_r$ being respectively the Young modulus and Poisson ratio of rust. The bulk modulus is the reciprocal value of the compressibility of rust, which is under isothermal conditions  expressed as 
\begin{equation}\label{compressibility}
\dfrac{1}{K_r} = -\dfrac{1}{V_{r,d}}\left(\dfrac{\partial V_{r,d}}{\partial p}\right)
\end{equation}  
Integration of (\ref{compressibility}) leads to 
\beq 
V_{r,d} = V_0\,\exp\left(-\frac{p}{K_r}\right)
\eeq 
where $V_0$ is an integration constant, which has the meaning
of the volume that would be occupied by the rust at zero pressure. By inversion, it is possible to express the pressure as
\begin{equation}\label{compressibility5}
p = K_{r} \ln \dfrac{V_0}{V_{r,d}} 
\end{equation}    
The rust layer of thickness $ t_{r} =t_{cor}+u_c$ occupies volume $ V_{r,d} = \pi((a+u_{c})^2-(a-t_{cor})^2)L $. The corrosion penetration is usually much smaller than the bar diameter. The assumption that $ a \gg t_{r} $ leads to $ V_{r,d} \approx 2\pi (t_{cor}+u_c) a L =2\pi t_{r} a L$.  Furthermore, the rust volume at zero pressure can be expressed as $ V_{0} = \pi((a+t_{cor}(\kappa-1))^2-(a-t_{cor})^2)L \approx 2\pi \kappa t_{cor} aL$, where 
\begin{equation}\label{expCoeff1}
\kappa = r_{h}\kappa_{h} + (1-r_{h})\kappa_{o}
\end{equation}  
is a volumetric expansion coefficient resulting from the disparity of molar volumes of steel and rust. This coefficient is evaluated from the mass fractions $ r_{h} $ and $1-r_{h}$ of iron hydroxy-oxides and iron oxides in rust produced at the given corrosion current density (see Fig. \ref{massRatRust}) and the corresponding molar volume ratios $\kappa_{o}$ and $\kappa_{h}$ of oxide rust/iron and hydroxy-oxide rust/iron, respectively. Substituting the expressions for $ V_{0} $ and $V_{r,d}$ into Eq. (\ref{compressibility5}) results in
\begin{equation}\label{compressibility4}
p = K_{r} \ln \left(\dfrac{\kappa t_{cor}}{t_{cor}+u_c}\right) 
\end{equation} 

Using the previously derived relation (\ref{displ4}), it is easy to eliminate the unknown pressure $p$ from Eq. (\ref{compressibility4}) and construct
an equation for a single unknown, $u_c$: 
\begin{equation}\label{subst1}
\dfrac{u_c}{C_c} = K_{r} \ln \left(\dfrac{\kappa t_{cor}}{t_{cor}+u_c}\right) 
\end{equation}    
This can be rewritten in the simple form
\begin{equation}\label{subst3}
G {\rm e}^{G} = F
\end{equation}  
in which $F = \kappa t_{cor}{\rm e}^{t_{cor}/(C_c K_{r})}/(C_c K_{r})  $ is a known right-hand side and the left-hand side
is a nonlinear function of unknown $G = (u_c + t_{cor})/(C_c K_{r})$. 
  For $F \geq 0$, Eq. (\ref{subst3})  has a unique real solution  $G=W(F)$, where $W$ is the so-called 
Lambert function, also known as the product logarithm. 
  Since $F$ is always positive for our problem, the solution of Eq. (\ref{subst1}) can be written as
\begin{equation}\label{subst4}
u_c = C_c K_{r}\, W\left(\dfrac{\kappa t_{cor}}{C_c K_{r}} \exp\left(\dfrac{t_{cor}}{C_c K_{r}}\right)\right) - t_{cor} 
\end{equation} 
The corresponding pressure $p$ is easily expressed as $ p =  u_c/C_c $. 

Let us note that even though the formula (\ref{subst4}) has been derived assuming axial symmetry, the real distribution of damage is not uniform on the boundary between steel and concrete due to the gradual localisation of cracks. For this reason, the assumption of axisymmetry will be violated and the formula will be evaluated point-wise on the steel-concrete boundary. 

\FloatBarrier
\subsection{Precipitation eigenstrain}
\label{sec:mechanics_eigenstrain}

$\mathrm{Fe}^{2+}$ and $\mathrm{Fe}^{3+}$ ions transported into the pore space gradually precipitate there into iron oxides and hydroxy-oxides. As was the case for the dense rust layer adjacent to the rebar, rust accumulates in the confined conditions of concrete pore space and exerts pressure on concrete pore walls \cite{Angst2019a, Angst2018a} (see Fig. \ref{Fig2} for a graphical illustration). This statement is supported both by the similarity with the well-described damage mechanism of precipitating salts in porous materials \cite{Scherer1999, Flatt2014, Flatt2017a, Coussy2006, Castellazzi2013, Koniorczyk2012, Espinosa2008} and the recent simulations of \citet{Korec2023}, which revealed that this mechanism could considerably contribute to corrosion-induced fracture in its early stages until the pore space surrounding rebar is locally filled. Macroscopic stress resulting from the accumulation of rust in the pore space under confined conditions is simulated with the precipitation eigenstrain proposed by \citet{Korec2023}, based on the previous works of \citet{Coussy2006}, \citet{Krajcinovic1992} and \citet{mura1987}. The precipitation eigenstrain is calculated as 
\begin{equation}\label{crystEigStr2}
\bm{\varepsilon}_{\star} = C S_{r}\bm{1}, \text{ with } C = \dfrac{(1-\nu_{c})K_{r}\left(\kappa - 1 \right)}{(1+\nu_{c})K_{r}+(2-4\nu_{c})K}
\end{equation}
where  $ K = E/(3(1-2\nu_{c}))$ is the bulk modulus of rust-filled concrete calculated from Young's modulus of rust-filled concrete $E$ and concrete Poisson ratio $\nu_{c}$. Because the mechanical properties of rust and concrete are very different, $E = (1-\theta_{r})E_{c} + \theta_{r}E_{r}$ is interpolated between Young's modulus of rust-free concrete $E_c$ and rust Young modulus $E_r$ following the rule of mixtures.

\FloatBarrier
\subsection{Damage--dependent diffusivity tensor}
\label{sec:DamDepTen}

Dissolved $\mathrm{Fe}^{2+}$ and $\mathrm{Fe}^{3+}$ ions are transported much more easily through emerging cracks than through the surrounding concrete matrix. For this reason, the damage-dependent diffusivity tensor \cite{Korec2023,Wu2016} is considered, and defined as
\begin{equation}\label{diffusivity_tensor_2}
\theta_{l}\boldsymbol{D}_{\alpha}=\theta_{l}(1-\phi)D_{m,\alpha}\boldsymbol{1} + \phi D_{c,\alpha} \boldsymbol{1}, \quad \alpha = II, III
\end{equation}    
where $ D_{m,\alpha} $ is the diffusivity of the considered species in concrete and  $D_{c,\alpha} \gg D_{m,\alpha} $ controls the diffusivity of the cracked material.

\subsection{Overview of the governing equations}
\label{Sec:govEq}
The solved coupled problem comprises six differential equations (\ref{summGovEq1})--(\ref{summGovEq6}) for six unknown variables -- displacement vector $\mathbf{u}$, phase-field variable $\phi$, $\mathrm{Fe}^{2+}$ ions concentration $c_{II}$, $\mathrm{Fe}^{3+}$ ions concentration $c_{III}$, hydroxy-oxide rust volume fraction $\theta_{h}$ and oxide rust volume fraction $\theta_{o}$: 
\begin{subequations}\label{summGovEq}
\begin{align}
\label{summGovEq1}
\bm{\nabla} \cdot\left(g(\phi) \mathcal{\bm{C}}_{e}^{c}: \left(\bm{\nabla}_{s} \mathbf{u} - \bm{\varepsilon}_{\star}\right)\right)+\overline{\mathbf{b}} &=0 \,\,\,\,\,\, \text{ in } \Omega^{c}\\
\label{summGovEq2}
-\dfrac{1}{2}\frac{\dd g(\phi)}{\mathrm{~d} \phi} \mathcal{H}(t)+\dfrac{\ell}{\pi} G_{f} \nabla^{2} \phi-\dfrac{G_{f}}{\pi \ell}(1-\phi) &=0 \,\,\,\,\,\,\, \text{ in } \Omega^{c} \\
\label{summGovEq3}
\frac{\partial \left(\theta_{l}c_{II}\right)}{\partial t} - \bm{\nabla} \cdot \left(\theta_{l}\bm{D}_{II}\cdot\nabla c_{II} \right) &=  \theta_{l} R_{II} \,\,\,\,\,\, \text{ in } \Omega^{c} \\
\label{summGovEq4}
\frac{\partial \left(\theta_{l}c_{III}\right)}{\partial t} - \bm{\nabla} \cdot \left(\theta_{l}\bm{D}_{III}\cdot\nabla c_{III} \right) &=  \theta_{l} R_{III} \,\,\,\,\,\, \text{ in } \Omega^{c} \\
\label{summGovEq5}
\frac{\partial \theta_{h}}{\partial t} &=  \dfrac{M_{h}}{\rho_{h}} \theta_{l} R_{h} \,\,\,\,\,\, \text{ in } \Omega^{c} \\
\label{summGovEq6}
\frac{\partial \theta_{o}}{\partial t} &=  \dfrac{M_{o}}{\rho_{o}} \theta_{l} R_{o} \,\,\,\,\,\, \text{ in } \Omega^{c}
\end{align}
\end{subequations}
\begin{subequations}\label{summGovBC}
Differential equations (\ref{summGovEq1})--(\ref{summGovEq4}) are associated (in the presented order) with boundary conditions 
\begin{align} 
\label{summBC1}
\mathbf{u} &= \overline{\mathbf{u}} \,\,\,\,\,\,\, \text{ on } \,\,\Gamma^{u} \hspace{1cm} \bm{\sigma} \cdot \mathbf{n} = \overline{\mathbf{t}} \,\,\,\,\,\,\, \text{ on } \,\,\Gamma^{t}\\ 
\label{summBC2}
\nabla \phi \cdot \mathbf{n} &= 0 \,\,\,\,\,\,\, \text{ on } \,\,\Gamma^{c} \cup \Gamma^{s}  \\
\label{summBC3}
\mathbf{n} \cdot \left(\boldsymbol{D}_{II} \cdot \nabla c_{II} \right) &= 0\,\,\,\,\,\,\, \text{ on } \,\,\Gamma^{c}, \hspace{0.5cm} \hspace{0.5cm} \mathbf{n} \cdot \left(\boldsymbol{D}_{II} \cdot \nabla c_{II} \right) = \frac{i_{a}}{z_{a}F}\,\,\,\,\,\,\, \text{ on } \,\,\Gamma^{s} \\ 
\label{summBC4}
\mathbf{n} \cdot \left(\boldsymbol{D}_{III} \cdot \nabla c_{III} \right) &= 0\,\,\,\,\,\,\, \text{ on } \,\,\Gamma^{c}\cup\Gamma^{s}  
\end{align}
\end{subequations}
Boundary conditions for equations (\ref{summGovEq5}) and (\ref{summGovEq6}) are not required because these equations do not contain any spatial derivatives.   

The resulting coupled system of differential equations is numerically solved with the finite element method using COMSOL Multiphysics software. 
The domain $\Omega^{c} \cup \Omega^{s}$ is discretised with linear triangular elements and a staggered solution scheme is employed. To achieve mesh-independent results, $\ell = 3$ mm is chosen such that $\ell$ would be five times larger than the maximum element size \cite{Kristensen2021}. 

\FloatBarrier
\section{Results}
\label{Sec:Results}

After the presentation of parameter values employed for numerical simulations (Section \ref{Sec:modelParam}), the observed characteristic features of the model are discussed in Section \ref{Sec:GenAspMod}, together with the implications for the understanding of corrosion-induced cracking. Then, in Section \ref{Sec:ResValCrWidSlope}, impressed current tests with varying corrosion current density from the comprehensive experimental study of \citet{Pedrosa2017} are simulated. The predicted relations between the surface crack and the thickness of the corroded steel layer are fitted with linear functions and their slopes are compared with the experimental measurements, revealing the ability of the model to capture the decrease of the slope with increasing corrosion current density. Based on the obtained results, a correction factor to be applied on the crack width slope obtained from accelerated impressed current tests is proposed, allowing for extrapolation of results to naturally corroding structures.     
\subsection{Choice of model parameters}
\label{Sec:modelParam}

The mechanical parameters listed in Table \ref{tab:tableMechRust1} correspond to the concrete specimens employed in the comprehensive experimental study by \citet{Pedrosa2017}, in which the corrosion-induced cracking under different values of applied corrosion current density was analysed. The 15 x 15 x 50 cm samples reinforced with a 16 mm rebar were separated into two groups with curing times of 28 and 147 days, which led to different mechanical properties. \citet{Pedrosa2017} measured the tensile and compressive strength. The fracture energy is estimated using the formula of \citet{Bazant2002}. For the steel rebar, Young's modulus of 205 GPa and Poisson's ratio of 0.28 are assumed, as these are common values for steel. 

\begin{table}[htb!]
\begin{small}
\begin{longtable}[ht]{p{5cm} p{2.5cm} p{2cm} p{1.5cm}}\toprule
\multicolumn{4}{ c }{\textbf{Properties of concrete - 28 and 147 days cured samples}} \\
\toprule
\textbf{Parameter} & \textbf{Value} & \textbf{Unit} & \textbf{Source}\\
\toprule
\toprule
Compressive strength $f_{c,cube}$ & 37.5 \& 54.7 & MPa & \cite{Pedrosa2017}\\
\midrule
Tensile strength $f_{t}$ & 2.2 \& 3.9 & MPa &  \cite{Pedrosa2017} \\
\midrule
Young's modulus $E_{c}$ & 33 \& 36 & GPa & \cite{standard2004eurocode}  \\
\midrule
Poisson's ratio $\nu_{c}$ & 0.2 \& 0.2 & - & \cite{standard2004eurocode}  \\
\midrule
Fracture energy $G_{f}$ & 95 \& 114 & N m$^{-1}$ & \cite{Bazant2002} \\
\bottomrule
\caption{Model parameters: mechanical properties of concrete, based on the measurements by \citet{Pedrosa2017} and literature data.}
\label{tab:tableMechRust1}
\end{longtable}
\end{small}
\end{table}

Only the capillary porosity of cement paste is considered to be relevant in the vicinity of rebar, and it is estimated using the model of \citet{Powers1946}, assuming the degree of hydration of 0.9. The values of the initial diffusivity of $\mathrm{Fe^{2+}}$ ions and $\mathrm{Fe^{3+}}$ ions in concrete are adopted from the study of \citet{stefanoni_kinetic_2018}, where they were estimated from the known diffusivity in solution assuming that the ratio between the diffusivity in water and concrete is the same as for chloride ions. The diffusivity of iron ions in cracks was considered identical to the diffusivity in the pore solution. 

\begin{table}[htb!]
\begin{small}
\begin{longtable}{p{10cm} p{2cm} p{2cm} p{2cm}}
\toprule
\textbf{Parameter} & \textbf{Value} & \textbf{Unit} & \textbf{Source} \\
\toprule
\multicolumn{4}{ c }{\textbf{Properties of rust}} \\
\toprule
\toprule
Young's modulus $E_{r}$ & $ 500  $ & MPa & \cite{ZHAO201619} \\
\midrule
Poisson's ratio $\nu_{r}$ & $ 0.4  $ & - &  \cite{ZHAO201619} \\
\midrule
Molar volume ratio of iron oxide rust to iron $\kappa_{o}$ & $ 2 $ & - & \cite{ZHAO201619,Vu2005,Zhang2019c} \\
\midrule
Molar volume ratio of iron hydroxy-oxide rust to iron $\kappa_{h}$ & $ 3.3 $ & - & \cite{ZHAO201619,Vu2005,Zhang2019c} \\
\midrule
Diffusivity of iron ions in rust $D_{r}$ & $ 10^{-10} $ & m$^{2}$ s$^{-1}$ & \cite{Ansari2019}  \\
\toprule
\multicolumn{4}{ c }{\textbf{Transport properties of concrete}} \\
\toprule
\toprule
Capillary porosity $p_{0}$ & $ 0.26 $ & - & \cite{Powers1946} \\
\midrule
Initial concrete diffusivity $\theta_{l}D_{m,II}$ and $\theta_{l}D_{m,III}$ & $ 10^{-11} $ & m$^{2}$ s$^{-1}$ & \cite{stefanoni_kinetic_2018}  \\
\midrule
Cracked concrete diffusivity $ D_{c,II} $ and $ D_{c,III} $ & $ 7\cdot10^{-10}  $ & m$^{2}$ s$^{-1}$ &  \cite{stefanoni_kinetic_2018,Leupin2021} \\
\bottomrule
\caption{Model parameters: properties of rust and the transport properties of concrete.}
\label{tab:tableMechRust2}
\end{longtable}
\end{small}
\end{table}

Molar volume ratios of iron oxide rust/iron and hydroxy-oxide rust/iron characterising rust volumetric expansion are chosen in the range of values reported by \citet{ZHAO201619,Vu2005} and \citet{Zhang2019c} for commonly present iron oxides and iron hydroxy-oxides. The measurements of Young's modulus and Poisson's ratio of rust are infamously scattered in the current literature, so two intermediate values within the range reported by \citet{ZHAO201619} are considered; they were identified to lead to an accurate prediction of crack width in the previous study of the authors \cite{Korec2023}. The diffusivity for iron ions in rust was adopted from the study of \citet{Ansari2019}. 

Let us also stress that in numerical simulations conducted within this study, uniform corrosion and thus the constant value of corrosion current density $ i_{a} $ are considered. This is because impressed current tests typically ensure uniform corrosion by artificially introducing high chloride concentration into concrete, as was done in tests by \citet{Pedrosa2017} considered in this study. For the results on the simulation of a non-uniform chloride-induced corrosion under natural conditions with a similar model as in this study, see Ref. \cite{Korec2024}.

\FloatBarrier
\subsection{General aspects of the model (the impact of the dense rust layer)}
\label{Sec:GenAspMod}

\begin{figure}[!htb]
    \centering
    \begin{subfigure}[!htb]{0.49\textwidth}
    \centering
    \includegraphics[width=\textwidth]{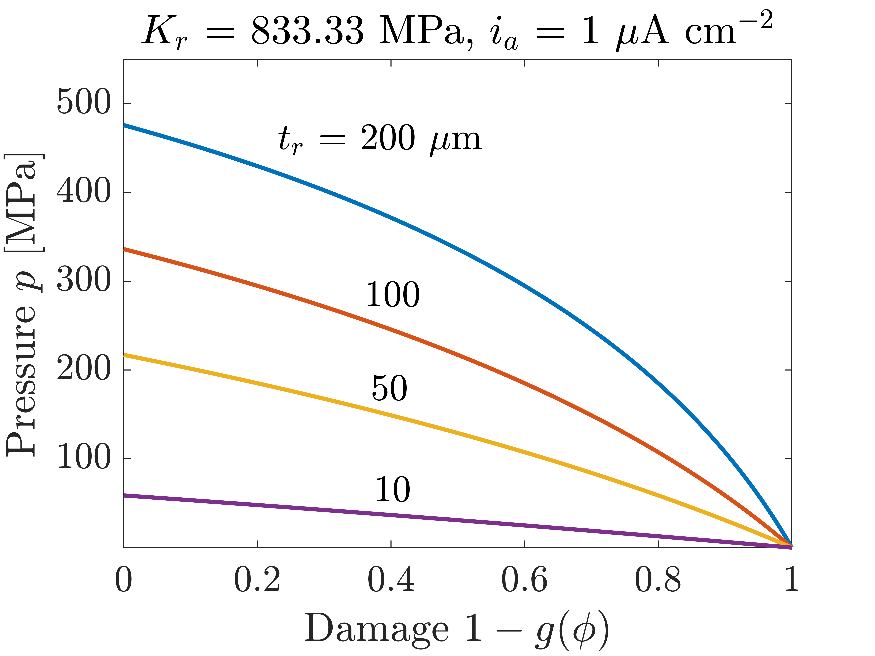}
    \caption{}
    \label{denLayPres1}    
    \end{subfigure}   
    \hfill 
    \centering
    \begin{subfigure}[!htb]{0.49\textwidth}
    \centering
    \includegraphics[width=\textwidth]{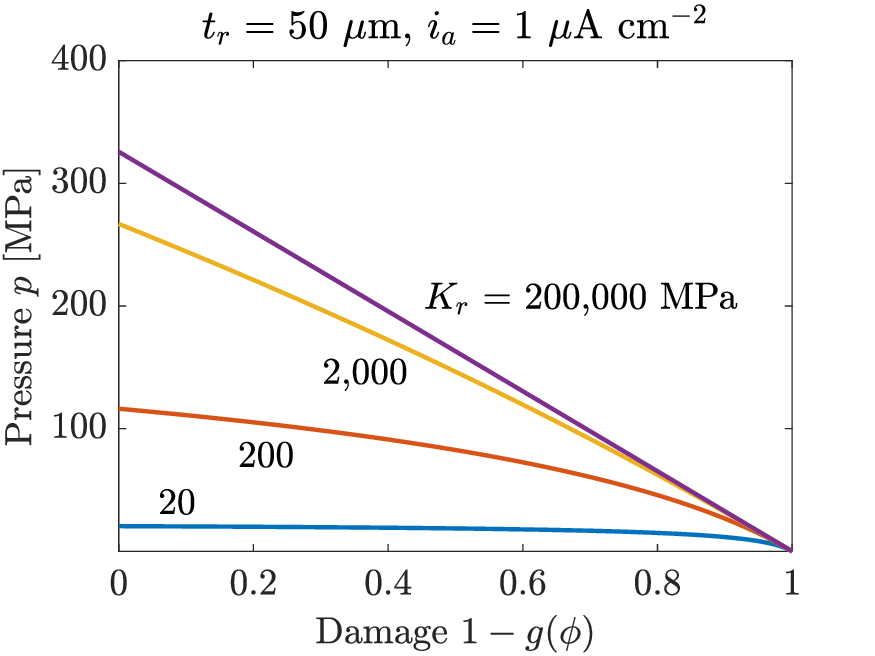}
    \caption{}
    \label{denLayPres2}    
    \end{subfigure} 
    \hfill 
    \centering
    \begin{subfigure}[!htb]{0.49\textwidth}
    \centering
    \includegraphics[width=\textwidth]{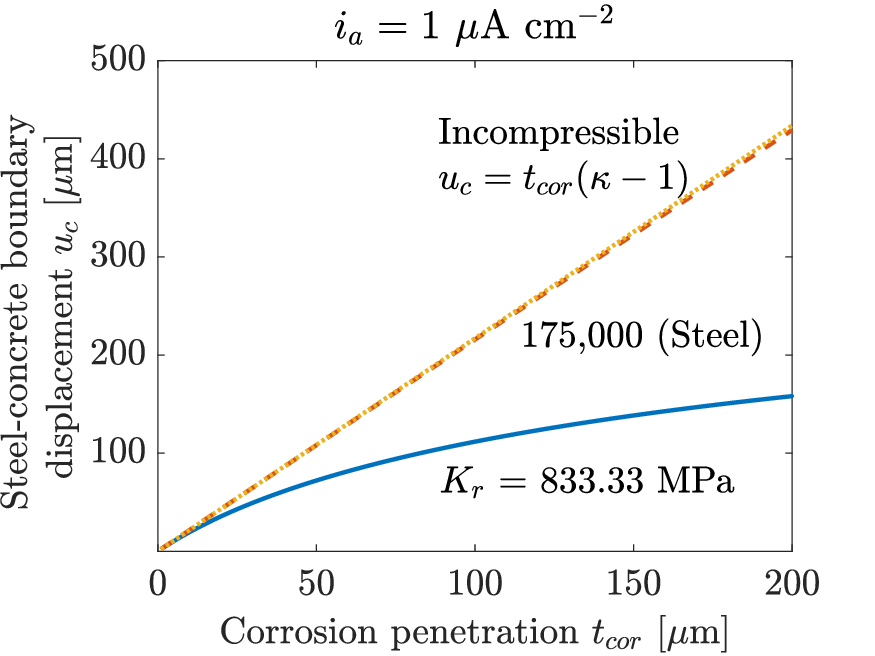}
    \caption{}
    \label{denLayPres3}    
    \end{subfigure}    
\caption{Pressure $p$ exerted by a dense rust layer on concrete -- (a) Pressure $p$ increases with the thickness of the corroded steel layer but decreases when concrete becomes more compliant due to damage. In many currently available models, rust is often simplified to be incompressible. However, the stiffness of rust has a profound impact on pressure predictions (b). Results in (c) reveal that the assumption of rust incompressibility leads to nearly identical results as if rust had the same elastic properties as steel.}
\label{fig:genAspMod}
\end{figure}

\begin{figure}[!htb]    
    \centering
    \begin{subfigure}[!htb]{0.49\textwidth}
    \centering
    \includegraphics[width=\textwidth]{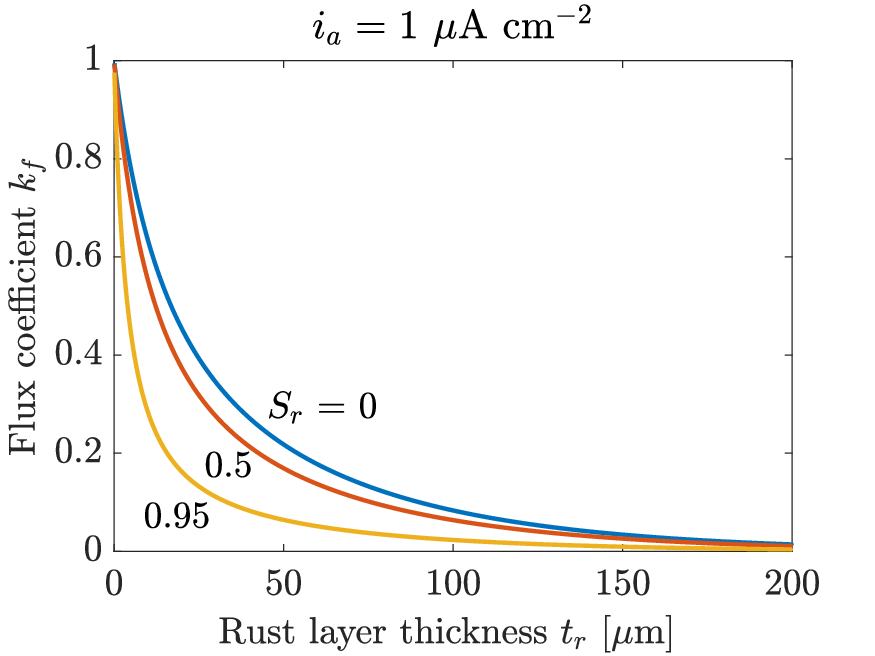}
    \caption{}
    \label{fig:fluxCoeff}    
    \end{subfigure}   
    \hfill 
    \centering
    \begin{subfigure}[!htb]{0.49\textwidth}
    \centering
    \includegraphics[width=\textwidth]{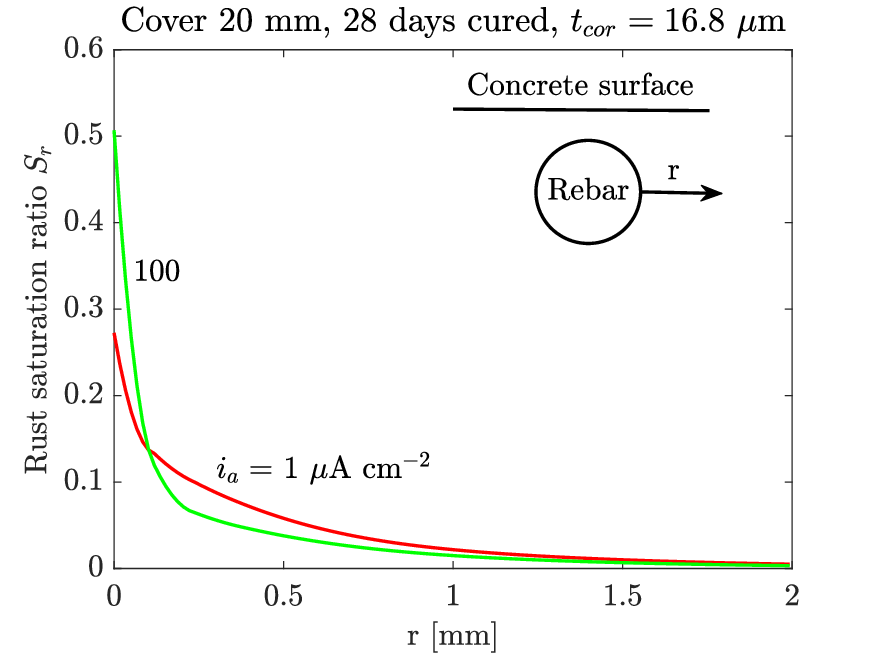}
    \caption{}
    \label{fig:SpCompar}    
    \end{subfigure}   
\caption{The impact of rust layer thickness and concrete rust saturation on the flux reduction coefficient $k_{f}$ and the impact of the magnitude of corrosion current on the distribution of rust in pore space -- (a) Flux reduction coefficient $k_{f}\in [0,1]$ reduces the Faraday law dictated flux $J_{II,Far}$, such that the flux released to concrete pore space is $J_{II} = k_{f}J_{II,Far}$. This reduction is the result of the deposition of a part of the precipitates in a dense rust layer adjacent to the corroding steel surface. The coefficient $k_{f}$ decreases with the thickness of the dense rust layer and with the saturation of the pore space by precipitates, such that $J_{II}$ eventually vanishes and all released $\mathrm{Fe}^{2+}$ ions contribute to the formation of the dense rust layer. (b) For high corrosion current densities in accelerated tests, rust accumulates closer to the steel surface than during corrosion in natural conditions, and thus the maximum rust saturation of pores is higher.}
\label{fig:genAspMod2}
\end{figure}

In the proposed model, rust precipitates either in the dense rust layer adjacent to the corroding rebar surface or deeper in the pore space of concrete. The flux of $\mathrm{Fe}^{2+}$ ions released from the corroding steel surface according to Faraday's law is divided between these two processes based on the flux reduction coefficient $k_{f}\in \left[ 0,1 \right]$. As can be seen in Fig. \ref{fig:fluxCoeff}, $ k_{f} = 1 $ if there is no dense rust layer and all released ferrous ions are transported into the concrete pore space. With the advance of the corrosion process, the thickness of the dense rust layer increases, decreasing the portion of $\mathrm{Fe}^{2+}$ ion flux that is released to the concrete pore space.
For a thick rust layer, $ J_{II} $ eventually vanishes, leaving all released $\mathrm{Fe}^{2+}$ ions to contribute to the build-up of the dense rust layer. Also, as saturation of the pore space with precipitates increases, the $\mathrm{Fe}^{2+}$ ion diffusivity in concrete is reduced, which decreases the ability of $\mathrm{Fe}^{2+}$ ions to escape the dense rust layer.

The accumulation of precipitated rust in (A) a dense layer of corrosion products in the space vacated by steel corrosion, and (B) the concrete pore space under constrained conditions, induces pressure on the concrete matrix. Pressure $ p $ in the dense rust layer increases with its thickness $t_{cor}$ (see Figs.~\ref{denLayPres1} and \ref{denLayPres3}). However, it also non-linearly decreases with damage, since concrete becomes more compliant and the expansion of rust is then less constrained (Figs.~\ref{denLayPres1} and \ref{denLayPres2}). The mechanical properties of rust have a strong influence on $p$ (see the dependence on the bulk modulus of rust in Fig. \ref{denLayPres2}). If the bulk modulus of rust was similar to steel, pressure $p$ would decrease almost in proportion to the degradation factor $g(\phi)$. However, for lower bulk moduli of rust, in tens or hundreds of MPa, the decrease deviates from linearity and the induced pressure $p$ is significantly smaller.  Many currently available models neglect the impact of rust compressibility and simplify the steel-concrete boundary displacement to be $ u_{c} = t_{cor}(\kappa - 1) $ as if rust was incompressible. As demonstrated in Fig. \ref{denLayPres3}, this assumption leads to a nearly equivalent displacement prediction as if rust elastic properties were equivalent to those of steel. Since the experimental measurement of Young's modulus and Poisson's ratio of rust are quite scattered \cite{ZHAO201619} in the range from 100 MPa to 360 GPa, our results stress the need for accurate characterization studies.

From the conducted simulations, it was observed that the distribution of rust in pore space differed for natural-like low corrosion current densities (about 1 \unit{\micro\ampere\per\centi\metre^2} and lower) compared to high current densities (tens or hundreds of \unit{\micro\ampere\per\centi\metre^2}) typical for corrosion induced by accelerated impressed current (see Fig. \ref{fig:SpCompar}). In the case of high current densities, the rust was distributed closer to the steel surface than for low current densities, which led to a significantly higher rust saturation of pores in the vicinity of steel rebar. Thus, pores are blocked with rust earlier than in natural conditions, not allowing for additional transport of dissolved iron species into pore space. This suggests that during the natural corrosion process, significantly more dissolved iron species can escape to the concrete pore space and precipitate there. The rapid increase of the rust saturation for high current densities probably results from significantly higher concentrations of released $\mathrm{Fe}^{2+}$ ions, which accelerates the precipitation reactions (\ref{reac3}) and (\ref{reac4}). Similar results were obtained in the computational study of \citet{stefanoni_kinetic_2018}.          

\FloatBarrier
\subsection{The impact of the magnitude of applied corrosion current density on crack width---analysis and validation}
\label{Sec:ResValCrWidSlope}

\begin{figure}[!htb]
    \centering
    \begin{subfigure}[!htb]{0.49\textwidth}
    \centering
    \includegraphics[width=\textwidth]{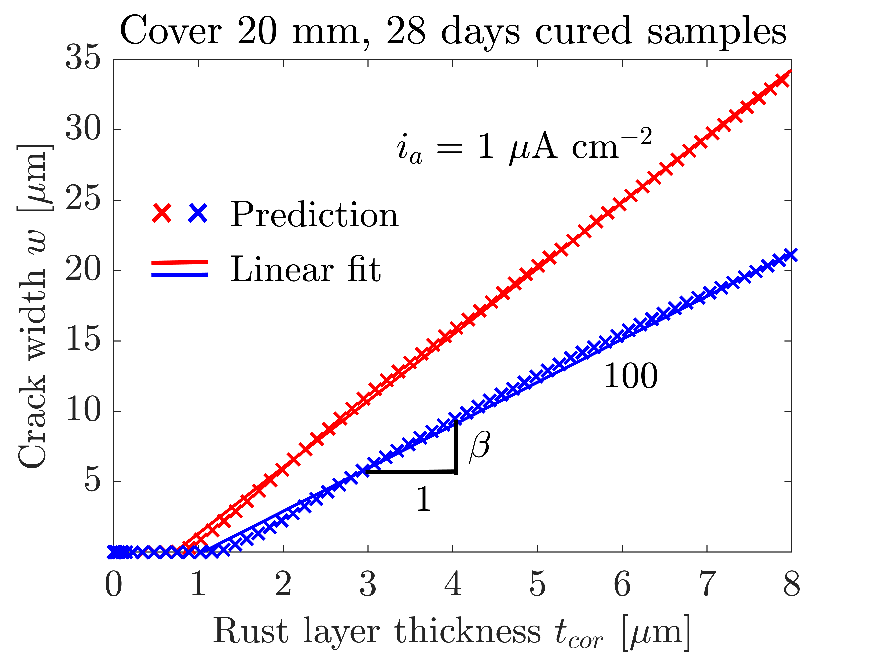}
    \caption{}
    \label{FigCrWidCover20}    
    \end{subfigure}          
    \hfill 
    \centering
    \begin{subfigure}[!htb]{0.49\textwidth}
    \centering
    \includegraphics[width=\textwidth]{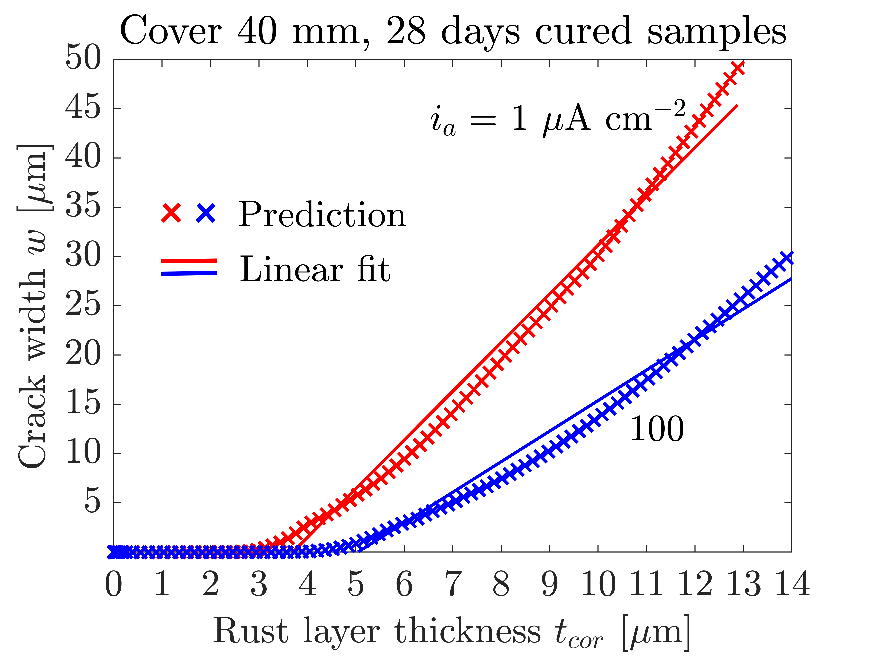}
    \caption{}
    \label{FigCrWidCover40}    
    \end{subfigure} 
    \centering
    \begin{subfigure}[!htb]{0.49\textwidth}
    \centering
    \includegraphics[width=\textwidth]{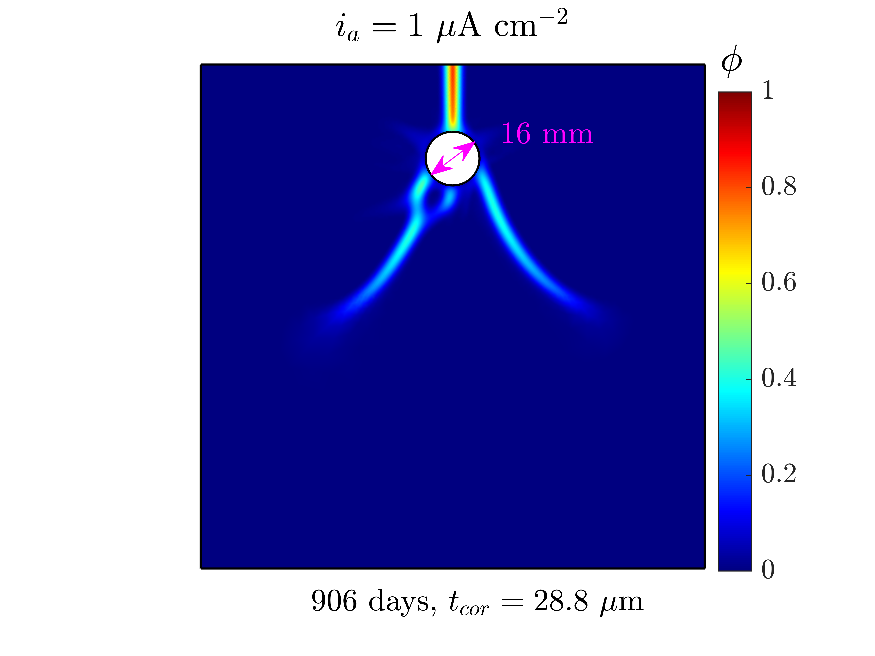}
    \caption{}
    \label{FigPFdist}    
    \end{subfigure}         
\caption{Growth of the upper surface crack width $w$ depending on the thickness of the corroded steel layer $t_{cor}$ for three values of applied corrosion current density and for concrete cover (a) 20 mm and (b) 40 mm. For identical $t_{cor}$, $w$ decreases with the increasing applied current density as the result of the changing chemical composition of rust, causing its increasing density and thus decreased corrosion-induced pressure. The predicted values in the range where $w \geq 0.5$ \unit{\micro\metre} are linearly fitted by $w = \beta t_{cor} + \zeta $. While $w(t_{cor})$ is nearly linear for the smaller concrete cover of 20 mm, non-linear dependence is observed for the larger concrete cover of 40 mm. In (c), the typical fracture patterns characterised by the contours of phase-field variable $\phi$ are depicted for the simulated impressed current test conducted on 28 days old concrete with a cover of 20 mm and the corrosion current of $i_{a} = 1$ \unit{\micro\ampere\per\centi\metre^2}.}
\label{fig:CrWidth}
\end{figure}

Fig. \ref{FigPFdist} shows the geometry and the typical fracture pattern observed for the simulated impressed current tests. The main crack propagates across the shortest distance between the concrete surface and the steel rebar, perpendicularly to the concrete surface, and opens wide with the ongoing corrosion process. In addition, two other lateral cracks develop. The surface crack width 
\begin{equation}
w = \int_{\Gamma^{us}} (\bm{\varepsilon}_{d})_{x} \dd \Gamma =  \int_{\Gamma^{us}} (1-g(\phi))(\bm{\varepsilon}_{x}-(\bm{\varepsilon}_{\star})_{x}) \dd \Gamma
\end{equation}    
is calculated as the integral of the $x$-component (i.e., the normal component in the direction perpendicular to the crack) of the inelastic strain tensor $ \bm{\varepsilon}_{d}  = \bm{\varepsilon} - \bm{\varepsilon}_{e} - \bm{\varepsilon}_{\star} $  over the upper concrete surface \citep{Navidtehrani2022}.
The inelastic strain is obtained by subtracting the eigenstrain $\bm{\varepsilon}_{\star}$ and the elastic strain $\bm{\varepsilon}_{e} = (\mathcal{\bm{C}}_{e}^{c})^{-1}:\bm{\sigma}=g(\phi)(\bm{\varepsilon}-\bm{\varepsilon}_{\star})$ from the total strain. For the same thickness of corroded steel layer $ t_{cor} $, the crack width decreases with increasing applied corrosion current density. This trend agrees with the experimental observations of \cite{Alonso1996,Pedrosa2017,Mangat1999,Vu2005,Vu2005a,ElMaaddawy2003,Mullard2011} and is the result of the hypothesis that the chemical composition of rust, in particular the mass ratio of iron oxides and iron hydroxy-oxides, changes with the magnitude of the applied current density (see Fig.~\ref{massRatRust}) such that the rust density increases with increasing current density, reducing the pressure caused by constrained expansion.   

The predicted crack width is linearly fitted by $w = \beta t_{cor} + \zeta $, so that the crack width slope $ \beta $ can be compared with its experimental counterpart measured by \citet{Pedrosa2017}, who employed the same fitting procedure on their experimentally measured crack widths. To prevent the distortion caused by nonlinearity for zero and very small crack widths, only $w \geq 1$ \unit{\micro\metre} was considered for fitting. Though linear fitting approximates the results very accurately for the thinner concrete of 20 mm (Fig. \ref{FigCrWidCover20}), the crack width increases superlinearly with $ t_{cor} $ for the thicker cover of 40 mm, which hinders the accuracy of the linear fitting procedure. For thicker concrete covers, the predicted crack width slope $ \beta $ is thus affected by the maximum evaluated $ t_{cor} $ and linear fitting is not optimal. 
In the discussed simulations, the maximal $ t_{cor} $ was 8 \unit{\micro\metre} and 12 \unit{\micro\metre} for the cases of 20 and 40 mm thick concrete cover, respectively.    

\begin{figure}[!htb]
    \begin{adjustbox}{minipage=\linewidth,scale=1}
    \centering
    \includegraphics[width=\textwidth]{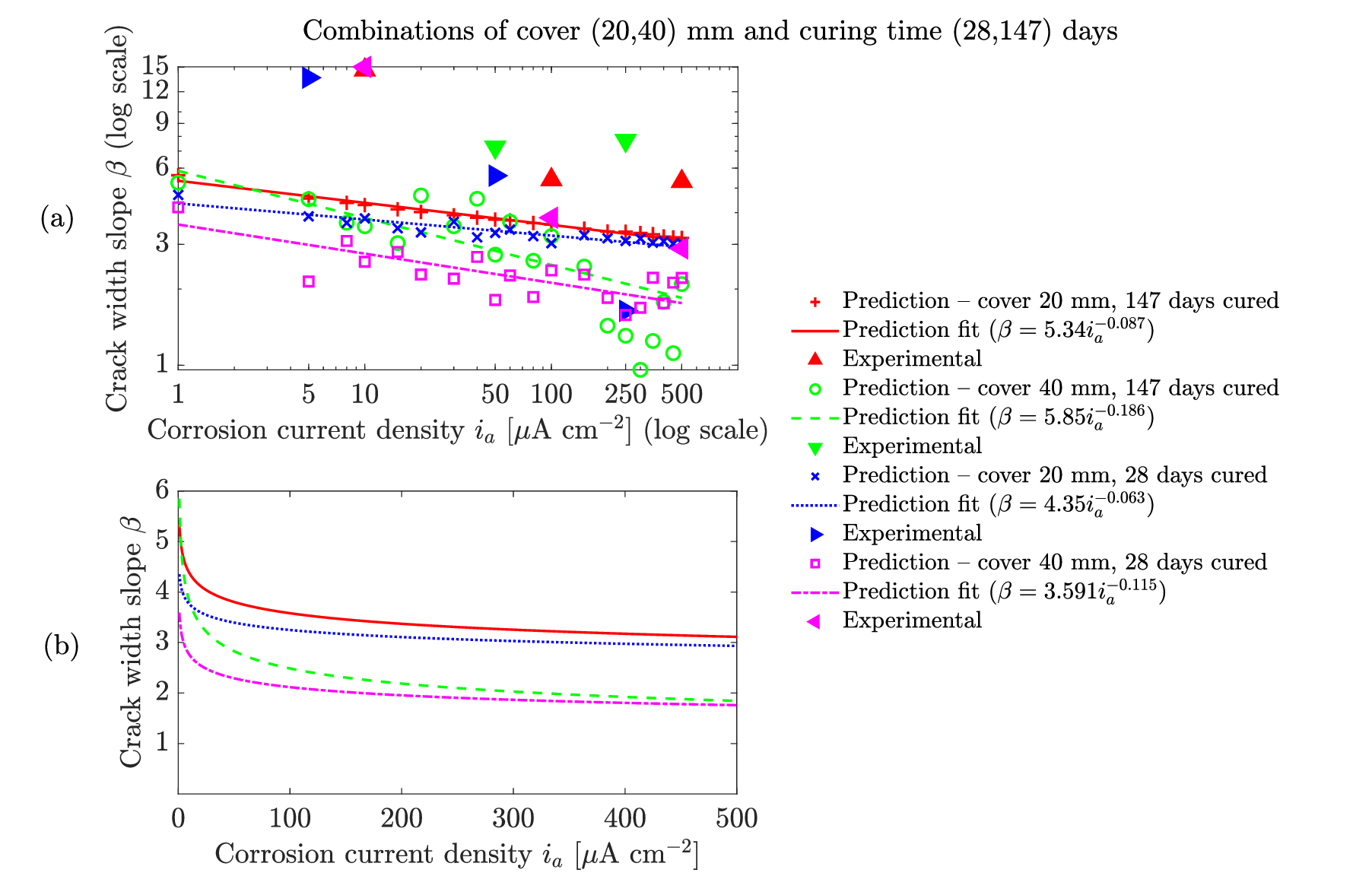}   
    \end{adjustbox} 
\caption{The slope $\beta$ of linearly fitted predicted crack width ($w = \beta t_{cor} + c $) is decreasing with the increasing applied corrosion current density. The reason is that the magnitude of applied current affects the composition of rust such that its density increases with the increasing applied current density, decreasing corrosion-induced pressure in the process. The comparison with 
experimental measurements (a) in log-log scale reveals that even though the decreasing trend of crack width slope with current density is captured well, experiments suggest even higher slopes for lower current densities such as $i_{a} = 5$ \unit{\micro\ampere\per\centi\metre^2}. Figure (b) allows to more easily observe that crack width slope rapidly decreases with increasing current density up to approximately $i_{a} = 50$ \unit{\micro\ampere\per\centi\metre^2}. From this, a more moderate decrease was observed.}     
\label{fig:CrWidSlope}
\end{figure}

The comparison of the predicted crack width slope $ \beta $ with the results recovered from the fitting of experimental measurements \cite{Pedrosa2017} reveals that the predicted data lie in the experimental range and that the trend characterised by the decay with a negative power of current density \cite{Pedrosa2017} is reproduced very well. For a better understanding of how well the various sets of predicted data agree with this trend, the predicted results are fitted with $ \beta = \gamma_{1} (i_{a}/i_{a,ref})^{\gamma_{2}} $, where $i_{a,ref} = 1$ \unit{\micro\ampere\per\centi\metre^2} is the reference corrosion current density. In Fig. \ref{CrWidSlope}, an excellent agreement with the data for the smaller concrete cover of 20 mm can be observed. On the other hand, the predictions for the thicker concrete cover of 40 mm are significantly more scattered, even though the overall trend still agrees. Fig. \ref{FigCrWidCover20} and Fig. \ref{FigCrWidCover40} suggest that the accuracy of the predictions is clearly related to how well can the data be fitted with a linear function. While data for the 20 mm cover are fitted very well, the non-linear dependency for the thicker 40 mm cover diminishes the representative value of $ \beta $. Also, it was observed that with the increasing thickness of the concrete cover, the model becomes numerically more sensitive such that there are several similar crack patterns that lead to a slightly different evolution of the surface crack width. The realization of a particular crack pattern appears to be triggered by small numerical differences. 

Critically, the proposed model captures the initially rapid decrease of $ \beta $ with increasing current density $i_{a}$ suggested by the experimental data of \citet{Pedrosa2017}. For current densities larger than approximately 50 \unit{\micro\ampere\per\centi\metre^2}, the decrease of $ \beta $ is more moderate. As can be seen in Fig. \ref{CrWidSlope}, this agrees qualitatively very well with the experiments. However, the magnitude of slopes observed by \citet{Pedrosa2017} for current densities smaller than 10 \unit{\micro\ampere\per\centi\metre^2} were even several times larger than those recovered from numerical simulations. Though the paper of \citet{Pedrosa2017} is currently the most comprehensive experimental study on the impact of the magnitude of impressed current on the crack width slope, the total number of tested specimen was arguably small (11 in total and every test was conducted only once which does not allow to evaluate experimental error). Thus, it is not clear whether the observed discrepancy between the quantitative values of numerical predictions and experiments can be attributed to the experimental error or some other phenomena unaccounted by the model. Clearly, more comprehensive experimental campaigns are needed. For current densities larger than 10 \unit{\micro\ampere\per\centi\metre^2}, numerical predictions agree quantitatively better with experimental measurements, though they mostly underestimate the experimental values.

\begin{figure}[!htb]
    \begin{adjustbox}{minipage=\linewidth,scale=1}
    \centering
    \begin{subfigure}[!htb]{0.49\textwidth}
    \includegraphics[width=\textwidth]{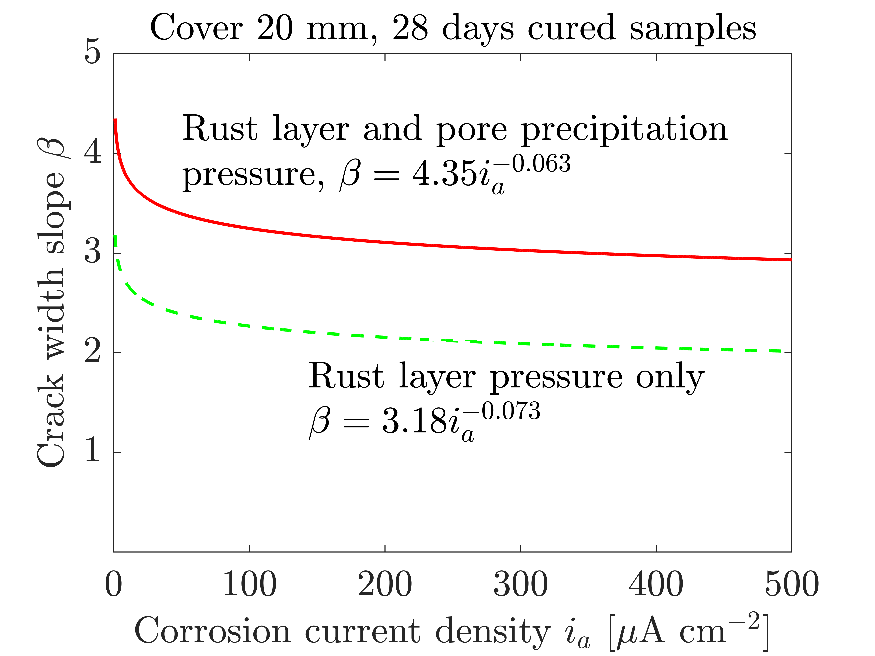}
    \caption{}
    \label{RusLayInfl}    
    \end{subfigure} 
    \hfill 
    \centering
    \begin{subfigure}[!htb]{0.49\textwidth}
    \centering
    \includegraphics[width=\textwidth]{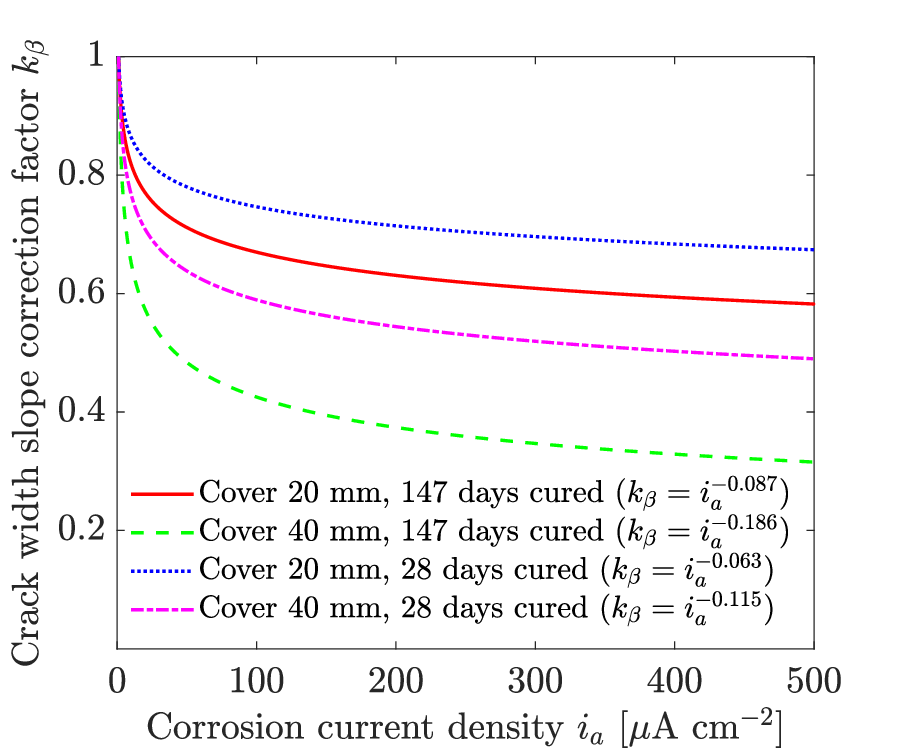}
    \caption{}
    \label{CorrFcator}    
    \end{subfigure}   
    \end{adjustbox} 
\caption{(a) The pressure of the dense rust layer accumulating in the volume vacated by steel corrosion is responsible for a considerable part of corrosion-induced pressure but the pressure resulting from the constrained accumulation of rust in pore space is not negligible. (b) The correction factor $ k_{\beta} $ represents the relative error caused by the current acceleration compared to the same test corroding under natural conditions. $ k_{\beta} $ decreases with the applied corrosion current density and the thickness of the concrete cover.}     
\label{fig:CrWidSlope2}
\end{figure}

The comparison of data for the four combinations of two curing times (leading to different mechanical properties) and two concrete covers in Fig. \ref{CrWidSlope} indicate that $ \beta $ increases with increasing curing time (i.e. enhanced mechanical properties). Also, a larger cover can lead to larger $ \beta $ for small natural-like current densities (within the evaluated case studies, this held for the combination of 40 mm cover and 147 days curing time), as was previously observed in Ref. \cite{Korec2023}. The modelling results in Fig. \ref{CrWidSlope} indicate that the influence of concrete cover is significantly more important than of curing time (i.e. concrete strength). Also, it appears that the larger the current density, the smaller the impact of curing time (i.e. concrete strength) on the crack width slope. The pressure generated by the constrained accumulation of the dense rust layer has a considerable impact on corrosion-induced fracture (see Fig \ref{RusLayInfl}), although the role of the pressure of rust precipitating in the pore space of concrete is also not negligible. However, one has to bear in mind that the proposed model overestimates the corrosion-induced pressure of the dense rust layer, as the entire volume of steel vacated by corrosion is assumed to be filled with rust regardless of the portion of ferrous ions escaping into pore space. In reality, the pressure of the dense rust layer can be reasonably expected to be smaller, raising the relative impact of the pressure of precipitates accumulating in the pore space.  

In Fig. \ref{CorrFcator}, the relative error of slope $ \beta $, caused by the acceleration of the applied current compared to corrosion in natural conditions, is evaluated in terms of newly proposed crack width slope correction factor $k_{\beta}$ inspired by the loading correction factor of \citet{Vu2005} for time-to-cracking. The correction factor $k_{\beta}$ was calculated as $ k_{\beta} = (i_{a}/i_{a,ref})^{\gamma_{2}} $. 
The factor $k_{\beta}$ can thus be interpreted as the relative error caused by the acceleration of the impressed current tests compared to results that could be expected in natural conditions when $i_{a} \approx 1$ \unit{\micro\ampere\per\centi\metre^2}. One can see that the error induced by current acceleration strongly increases with the thickness of the concrete cover. While for 147 days cured samples and $i_{a} = 500$ \unit{\micro\ampere\per\centi\metre^2}, $k_{\beta} = 0.58$ for the 20 mm thick cover, it drops to $k_{\beta} = 0.32$ for a 40 mm thick cover. The same trend was experimentally documented by \citet{Alonso1996} who reported even six times larger $ \beta $ if current density was accelerated from 10 to 100 \unit{\micro\ampere\per\centi\metre^2} for the concrete cover of 70 mm. Also, it can be observed that the impact of curing time on $k_{\beta}$ is smaller than the impact of the thickness of the concrete cover. 

The knowledge of the accurate slope of crack width with respect to corrosion penetration in natural conditions is important for the accurate predictions of the length of the corrosion propagation period (time since corrosion initiation to critical corrosion-induced delamination/spalling of concrete cover). Currently, durability estimates in national design codes are based on time until corrosion initiation, which is dictated by the time needed for the diffusion of aggressive species such as chlorides from exposed concrete surfaces through concrete cover to steel rebars. This approach is overly conservative, as even though corrosion initiation is usually the longest part of the corrosion process, the length of the propagation period can be substantial. For example, the results for a 40 mm thick concrete cover and concrete cured for 28 days indicate that under natural conditions (typically $i_{a} \leq 1$ \unit{\micro\ampere\per\centi\metre^2}) the propagation period would last for at least 7 years until the surface crack width of 0.3 mm (a commonly used Eurocode criterion of limit serviceability state) was reached. That said, even the crack opening of 0.3 mm can underestimate the time when serious delamination/spalling occurs \cite{Alonso1996,Andrade1993}.     

\section{Conclusions}
\label{Sec:Conclusions}

A phase-field-based model for corrosion-induced cracking of reinforced concrete specimens subjected to uniform corrosion was presented. Based on the experimental findings (Ref.  \cite{Zhang2019c}), the hypothesis that the chemical composition of rust characterized by the mass ratio of its main components (iron oxides and iron hydroxy-oxides) changes with the applied corrosion current density was tested. This affects the density of rust, which increases with increasing current density, thereby reducing the corrosion-induced pressure in the process. The main findings can be summarised as follows:
\begin{itemize}
\item The simulation results strongly support the hypothesis of changing chemical composition of rust with the applied corrosion current density as the explanation of the experimentally observed slower crack growth (with respect to corrosion penetration) in accelerated impressed current tests compared to corrosion in natural conditions \cite{Alonso1996,Pedrosa2017,Mangat1999,Vu2005,Vu2005a,ElMaaddawy2003,Mullard2011,Vu2005}, which has been debated for nearly three decades. 
\item Simulations support the experimentally recovered conclusion of \citet{Pedrosa2017} that the decay of the slope of the crack width as a function of corrosion penetration is proportional to a negative power of corrosion current density, especially for thinner concrete covers.
\item The non-linearity of the crack width dependence on corrosion penetration increases with the thickness of the concrete cover. 
\item The relative error in the estimate of crack width slope compared to corrosion in natural conditions increases with the applied corrosion current density and the thickness of the concrete cover.  
\item Current design codes base the corrosion durability estimates of concrete structures on time for corrosion initiation (i.e. the transport of aggressive species such as chlorides through concrete cover). However, this approach is overly conservative and the consideration of the corrosion propagation stage of corrosion (since corrosion initiation to serious delamination/spalling) in the structure's service life can prolong the durability estimates for years. For a 40 mm thick concrete cover and concrete cured for 28 days, the proposed model predicts that under natural conditions ($i_{a} \leq 1$  \unit{\micro\ampere\per\centi\metre^2}) at least 7 years would pass until the surface crack width of 0.3 mm is reached.     
\end{itemize}  
A newly proposed crack width slope correction factor $k_{\beta}$ can be calculated from the proposed model and allows for the estimate of the relative error caused by the acceleration of the impressed current test to a specified value of corrosion current density. With this, the current density can be picked such that the error in the slope of crack width as a function of corrosion penetration remains (e.g.) below 10$\%$. In general, the proposed modelling framework allows for computational impressed current testing as the support of experimental efforts. 

\section{Acknowledgements}
\label{Acknowledge of funding}

The authors would like to express their gratitude to Prof Carmen Andrade (CIMNE International Center for Numerical Methods in Engineering, Barcelona) for her invaluable advice and stimulating discussions. E. Korec acknowledges financial support from the Imperial College President’s PhD Scholarships. M. Jirásek acknowledges previous financial support of the European Regional Development Fund (Center of Advanced Applied Sciences, project CZ.02.1.01/0.0/0.0/16\_19/0000778). E. Mart\'{\i}nez-Pa\~neda was supported by an UKRI Future Leaders Fellowship [grant MR/V024124/1]. We additionally acknowledge computational resources and support provided by the Imperial College Research Computing Service (http://doi.org/10.14469/hpc/2232). 

\appendix

\FloatBarrier
\section{Axisymmetric thick-walled concrete cylinder problem}
\label{Sec:A2}
\setcounter{figure}{0}
To calculate the pressure $ p $ of the dense rust layer in Section \ref{sec:mechanicsRustLay}, the concrete layer adjacent to the steel rebar is virtually isolated from the considered sample as a two-dimensional thick-walled cylinder with inner radius $ a $ (equal to the rebar radius) and outer radius $ b $ (see Fig. \ref{axiProblemFig}). 
\begin{figure}[!htb]
\begin{center}
\includegraphics[scale=0.85]{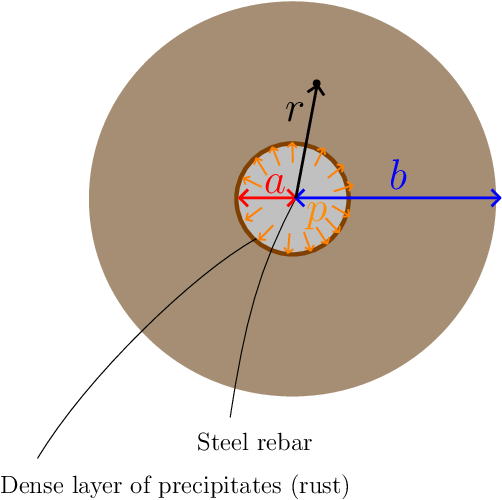}
\caption{The considered axisymmetric problem}
\label{axiProblemFig}
\end{center}
\end{figure}
Because the length of the cylinder $L$ is typically much larger than $b$ and material homogeneity and symmetric boundary conditions are assumed, the thick-walled cylinder can be mechanically described as an axisymmetric plane-strain problem. The plane is described by two polar coordinates---polar angle $ \varphi $ and radius $r$. The equilibrium equations for in-plane stress  components  read
\begin{equation}\label{equlibrium}
\begin{aligned}
&\frac{\partial \sigma_{r}}{\partial r}+\frac{1}{r} \frac{\partial \tau_{r \varphi}}{\partial \varphi}+\frac{1}{r}\left(\sigma_{r}-\sigma_{\varphi}\right)=0 \\
&\frac{\partial \tau_{r \varphi}}{\partial r}+\frac{1}{r} \frac{\partial \sigma_{\varphi }}{\partial \varphi}+\frac{2 \tau_{r \varphi}}{r}=0
\end{aligned}
\end{equation}
and the material behaviour is characterized by Hooke's law
for plane strain,
\begin{equation}\label{Hooke_law}
\begin{aligned}
&\sigma_{r}=\frac{E_{c}}{(1+\nu_{c})(1-2 \nu_{c})}\left(\nu_{c} \varepsilon_{\varphi}+(1-\nu_{c}) \varepsilon_{r}\right) \\
&\sigma_{\varphi}=\frac{E_{c}}{(1+\nu_{c})(1-2 \nu_{c})}\left(\nu_{c} \varepsilon_{r}+(1-\nu_{c}) \varepsilon_{\varphi}\right)\\
&\tau_{r\varphi}=\frac{E_c}{1+\nu_c}\varepsilon_{r\varphi}
\end{aligned}
\end{equation}
in which the strains are linked to the displacements by
\begin{equation}\label{strain_displacement}
\begin{aligned}
&\varepsilon_{r}=\frac{\partial u_{r}}{\partial r} \\
&\varepsilon_{\varphi}=\frac{1}{r} \frac{\partial u_{\varphi}}{\partial \varphi}+\frac{u_{r}}{r} \\
&\varepsilon_{r \varphi}=\frac{1}{2}\left(\frac{1}{r} \frac{\partial u_{r}}{\partial \varphi}+\frac{\partial u_{\varphi}}{\partial r}-\frac{u_{\varphi}}{r}\right)
\end{aligned}
\end{equation}

 The assumption of axial symmetry makes all considered quantities independent of $ \varphi $, and also leads to $ u_{\varphi} = 0 $. Therefore, from (\ref{strain_displacement}) follows that $ \varepsilon_{r\varphi} = 0$ and (\ref{Hooke_law}) yields $ \tau_{r\varphi} = 0 $. Equilibrium equations (\ref{equlibrium}) and strain-displacement equations (\ref{strain_displacement}) are simplified to
\begin{equation}\label{equlibrium_axi}
\begin{aligned}
&\frac{\partial \sigma_{r}}{\partial r}+\frac{1}{r}\left(\sigma_{r}-\sigma_{\varphi}\right)=0 
\end{aligned}
\end{equation}
\begin{equation}\label{strain_displacement_axi}
\begin{aligned}
&\varepsilon_{r}=\frac{\partial u_{r}}{\partial r} \\
&\varepsilon_{\varphi}=\frac{u_{r}}{r} 
\end{aligned}
\end{equation}
If (\ref{strain_displacement_axi}) is substituted into (\ref{Hooke_law}) and the resulting relations into (\ref{equlibrium_axi}), an ordinary differential equation is obtained
\begin{equation}
\frac{\dd^{2} u_r}{\dd r^{2}}+\frac{1}{r} \frac{\dd u_r}{\dd r}-\frac{1}{r^{2}} u_r=0
\end{equation}  
The general solution of this equation reads
\begin{equation}\label{axi_solution_disp}
u_r=C_{1} r+C_{2} \frac{1}{r}
\end{equation}
where $ C_{1} $ and $C_{2}$ are unknown real constants, to be calculated from boundary conditions. 
The corresponding strains and stresses are
\bea 
\varepsilon_r &=& C_1 - C_2\frac{1}{r^2} \\
\varepsilon_\varphi &=& C_{1} +C_{2} \frac{1}{r^2}\\
\label{A10}
\sigma_{r}&=&\frac{E_{c}}{(1+\nu_{c})(1-2 \nu_{c})}\left(C_1-C_2\frac{1-2\nu_c}{r^2}\right) \\
\sigma_{\varphi}&=&\frac{E_{c}}{(1+\nu_{c})(1-2 \nu_{c})}\left(C_1+C_2\frac{1-2\nu_c}{r^2}\right)
\eea 
Under uniform corrosion, the constrained volumetric expansion of the dense rust layer leads to pressure $p$ on the inner surface of the thick-walled cylinder, which generates
compressive stress $ \sigma_{r}(a) = -p $. It is supposed that the outer boundary of the concrete cylinder is stress-free, i.e., $ \sigma_{r}(b) = 0 $. Substituting these boundary
conditions into (\ref{A10}), two equations for the integration constants are obtained, from which
\bea 
C_1 &=&  \frac{p(1+\nu_c)(1-2\nu_c)}{E_c}\frac{a^2}{b^2-a^2}
\\
C_2 &=& \frac{p(1+\nu_c)}{E_c}\frac{a^2b^2}{b^2-a^2}
\eea 
The displacement at the inner boundary induced by pressure $p$
is then given by
\beq 
u_c=u_r(0) = C_1a+\frac{C_2}{a} = 
\frac{p(1+\nu_c)}{E_c}\frac{1}{b^2-a^2}\left((1-2\nu_c)a^3+ab^2\right)
\eeq 
Introducing a dimensionless shape factor $\alpha=b/a>1$,
the compliance is expressed as
\beq 
C_c=\frac{u_c}{p} = \frac{1+\nu_c}{E_c}\frac{\left(1-2\nu_c+\alpha^2\right)a}{\alpha^2-1}
\eeq 
and write the relation between pressure and displacement at the inner boundary in the form
\beq \label{eq:A16}
u_c = C_c\,p
\eeq 
The derivation has been done for concrete considered as an elastic material characterized by Young's modulus $E_c$
and Poisson's ratio $\nu_c$. If the concrete is damaged,
$E_c$ should be replaced by the secant modulus $g(\phi)E_{c}$
where $g$ is the degradation function and $\phi$ is the phase-field
variable that describes the state of damage. In general,
$\phi$ varies in the radial as well as circumferential direction, and the problem is no longer axially symmetric and does not have an analytical solution. For simplicity, 
the effect of damage is considered only approximately by using
a uniformly reduced modulus $E_{c,d} = g(\phi(a))E_{c}$, where the degradation function $g(\phi)$ is evaluated on the inner concrete boundary, i.e., at $r = a$. This means that the damage of concrete is simplified to be uniformly distributed over the thick-walled cylinder and the value of the damage is the same as on the inner concrete boundary. The final formula
for the compliance factor to be used in (\ref{eq:A16}) is then
\begin{equation}\label{displ5}
C_{c} = \dfrac{(1+\nu_{c})\left(\alpha^2 + 1-2 \nu_{c} \right)a}{E_{c,d}(\alpha^2-1)}
\end{equation}



\end{document}